\documentclass[a4paper,twocolumn,11pt,accepted=2026-03-10]{quantumarticle}
\pdfoutput=1
\usepackage{graphicx}
\usepackage{dcolumn}
\usepackage{bbm}
\usepackage{physics}
\usepackage{amssymb} 
\usepackage{csquotes}
\usepackage[title]{appendix}
\usepackage[normalem]{ulem}
\usepackage{comment}
\usepackage{bm}
\usepackage{subfigure}

\usepackage{csquotes}

\usepackage[colorlinks,citecolor=blue]{hyperref}
\usepackage{cleveref}

\usepackage[numbers,compress]{natbib}
\usepackage{comment}

\usepackage[dvipsnames]{xcolor}

\newcommand{\id}{\mathbb{I}}
\newcommand{\x}{\otimes}

\begin{document}


\title{A Hierarchy of Spectral Gap Certificates for Frustration-Free Spin Systems
}

\newcommand{\aqa}{$\langle aQa ^L\rangle $ Applied Quantum Algorithms, Universiteit Leiden}
\newcommand{\lorentz}{Instituut-Lorentz, Universiteit Leiden, Niels Bohrweg 2, 2333 CA Leiden, Netherlands}
\newcommand{\liacs}{LIACS, Universiteit Leiden, Niels Bohrweg 1, 2333 CA Leiden, Netherlands}
\newcommand{\univvienna}{University of Vienna, Faculty of Physics, Boltzmanngasse 5, A-1090 Vienna, Austria}
\newcommand{\mathvienna}{University of Vienna, Faculty of Mathematics,  Oskar-Morgenstern-Platz 1, A-1090 Vienna, Austria}
\newcommand{\padua}{Dipartimento di Fisica e Astronomia \enquote{G. Galilei}, Università di Padova, I-35131 Padova, Italy}
\newcommand{\qtech}{Padua Quantum Technologies Research Center, Università degli Studi di Padova, Italy I-35131, Padova, Italy}
\newcommand{\ulm}{Institute for Complex Quantum Systems, Ulm University, 89069 Ulm, Germany}
\newcommand{\iqst}{Center for Integrated Quantum Science and Technology (IQST), Ulm-Stuttgart, Germany}

\author{Kshiti Sneh Rai}
\thanks{These authors contributed equally.}
\affiliation{\aqa}
\affiliation{\lorentz}
\author{Ilya Kull}
\thanks{These authors contributed equally.}
\affiliation{\univvienna}
\author{Patrick Emonts}
\affiliation{\aqa}
\affiliation{\lorentz}
\affiliation{\ulm}
\affiliation{\iqst}
\author{Jordi Tura}
\affiliation{\aqa}
\affiliation{\lorentz}
\author{Norbert Schuch}
\affiliation{\univvienna}
\affiliation{\mathvienna}
\author{Flavio Baccari}
\affiliation{\padua}
\affiliation{\qtech}


\begin{abstract}

Estimating spectral gaps of quantum many-body Hamiltonians is a highly challenging computational task, even under assumptions of locality and translation-invariance. 
Yet, the quest for rigorous gap certificates is motivated by their broad applicability, ranging from many-body physics to quantum computing and classical sampling techniques.
Here we present a general method for obtaining lower bounds on the spectral gap of frustration-free quantum Hamiltonians in the thermodynamic limit.
We formulate the gap certification problem as a hierarchy of optimization problems (semidefinite programs) in which the certificate---a proof of a lower bound on the gap---is improved with increasing levels. 
Our approach   encompasses  existing finite-size methods, such as Knabe's bound and its subsequent improvements,  as  those    appear as particular possible solutions in our optimization, which is thus guaranteed to either match or surpass them.      
We demonstrate the power of the method on one-dimensional spin-chain models where we observe an improvement by several orders of magnitude over existing finite size criteria in both the accuracy of the lower bound on the gap, as well as the range of parameters in which a gap is detected.
\end{abstract}
\maketitle
\section{Introduction}
\label{sec:introduction}

A central question in many-body quantum physics is to determine whether a system is gapped or gapless in the thermodynamic limit. 
Indeed, the existence of a gap has important consequences for the physical properties of the low energy sector. 
For instance, it controls correlations and entanglement in the ground state, both in 1D~\cite{hastings_area_2007,hastings_spectral_2006} and in some 2D systems~\cite{anshu_area_2022}. 
Moreover, in 1D it directly relates to a notion of bounded complexity~\cite{landau_polynomial_2015}. 
Estimating the gap is also  crucial to several tasks, including the classification of quantum phases of matter~\cite{hastings_lieb-schultz-mattis_2004,bachmann_automorphic_2012}, adiabatic state preparation algorithms~\cite{ge_rapid_2016,rai_spectral_2024} and the identification of efficient Monte Carlo sampling methods~\cite{aharonov_adiabatic_2003,verstraete_criticality_2006}.

Proving the existence of a spectral gap is a highly non-trivial mathematical challenge.
Even for reasonable local Hamiltonians the task can be undecidable~\cite{cubitt_undecidability_2015,bausch_undecidability_2020,cubitt_undecidability_2022,perales-eceiza_undecidability_2024}.  
Several techniques to lower bound the spectral  gap have been developed for the better-behaved class of frustration-free Hamiltonians, i.e.\ systems in which the ground state minimizes each local term individually.
The most notable lower bounds are based on the \enquote{martingale} method~\cite{nachtergaele_spectral_1996} and finite size criteria~\cite{knabe_energy_1988,fannes_finitely_1992}.
More recently, methods inspired by the numerical  bootstrap approach to conformal field theory have been proposed to determine spectral properties of simple quantum Hamiltonians as well as quantum many body systems~\cite{berenstein_bootstrapping_2021, berenstein_semidefinite_2023, nancarrow_bootstrapping_2023}.

The martingale method is a powerful and widely applicable analytical technique.
In particular, using this approach, the question of the \textit{existence} of a gap has been solved for a wide class of for 1D frustration-free models~\cite{fannes_finitely_1992, nachtergaele_spectral_1996}. 
Such methods, powerful as they are, fall short in that they do not always allow to get an actual lower bound for the gap, 
and in cases in which  a number can be obtained, it is typically not a particularly tight lower bound.

Finite-size criteria, on the other hand, provide more realistic lower bounds on the spectral gap, with practical computational requirements.
Such methods establish a relation between the gap of the infinite system and a quantity pertaining to a finite-size system (with the same interaction). 
This approach was pioneered by Knabe who was able to provide a lower bound on the gap of the AKLT chain~\cite{knabe_energy_1988}.
Its subsequent improvements and generalizations~\cite{
gosset_local_2016,kastoryano_divide_2018, lemm_spectral_2019, anshu_improved_2020, lemm_quantitatively_2022}
have allowed to prove the gap in a wide range of systems relevant to both  many-body physics~\cite{
abdul-rahman_class_2020, guo_nonzero_2021, nachtergaele_spectral_2021, warze1_spectral_2022, lemm_gaplessness_2019, wouters_interrelations_2021,andrei_spin-one_2022}
and quantum information~\cite{haferkamp_improved_2021, lancien_correlation_2022}.
Despite recent advances~\cite{lemm_aklt_2019, pomata_demonstrating_2020, lemm_existence_2020}, 
crucial questions remain  open, particularly for systems in  spatial dimensions larger than one.

In this work we put forward a systematic way to exhaust the search  for gap certificates. 
Generalizing the  the core idea underlying both the martingale approach and finite-size criteria,
we   construct a hierarchy of   relaxations of the gap-estimation problem.
Each level $n$ in the hierarchy is a finite $n$-body optimization problem, the solution to which is a lower bound on the infinite-system gap.
As the level is increased  the quality of the lower bound is systematically improved.
Crucially, we are able to prove  that our method always matches or outperforms all finite-size criteria we know of, including the Knabe~\cite{knabe_energy_1988} and the Gosset--Mozgunov~\cite{ gosset_local_2016} bounds,
as those appear as  possible (but not necessarily optimal) solutions in our optimization problem.

We demonstrate the power of our method on several paradigmatic examples of 1D frustration-free models. 
Our method leads to the most accurate lower bound for the spin-1 AKLT chain do date, and detects the gap in parameter regimes far beyond those where  other finite-size criteria can detect.
While the presentation in the paper is specialized to  1D systems, the approach can be straightforwardly generalized to higher-dimensional lattices.

To put our results into context, 
we find it worthwhile to pause and briefly discuss the basic idea behind our method---and how it  generalizes   existing gap certification methods---on a slightly more  technical level.
The well known idea underlying all gap certification approaches is the 
fact that  a frustration-free Hamiltonian $H$ has a  gap greater or equal to $\delta$ if  the  operator $H^2 - \delta H$ is positive semidefinite (i.e.\ all its eigenvalues are non negative).
%
%
%
Existing  methods thus all amount to finding ways to decompose $H^2-\delta H$ as a sum of positive terms. 
Usually a decomposition is chosen  such that some terms are manifestly positive, while for others positivity can be proven given some positivity condition holds for an appropriately chosen local Hamiltonian---a finite size criterion. 
There are of course many possible choices of such decompositions, and finding the appropriate finite size criterion for a given system has thus far been an intricate task. 
The main idea of our method  is that the search for a positive decomposition of $H^2-\delta H$ can be automated and solved in an algorithmic way.
Namely we  formulate the gap estimation problem as
an optimization problem where $\delta $ is maximized under the constraint that $H^2 - \delta H$ can be written as a sum of positive terms.
A basic version of this idea has been proposed in Ref.~\cite{cruz_preparation_2022}.
Here we exhaust the search space in a systematic way through a hierarchy of optimization problems:  at the $n$-th level  of the hierarchy we optimize over all possible decompositions of $H^2 - \delta H$ as a sum of translation-invariant $n$-body geometrically local terms.

The rest of the paper is organized as follows:
In Section~\ref{sec:method}, we discuss the main technical ingredients of our method.
In Section~\ref{sec:existing_methods}, we show how previous finite-size criteria arise as special cases of the semidefinite program introduced in Section~\ref{sec:method}.
In Section~\ref{sec:numerics}, we present numerical results comparing lower bounds   obtained using our method against those obtained  with existing finite-size criteria.
\section{The method}
\label{sec:method}
We consider a spin chain comprised of particles defined on a $d-$dimensional Hilbert space $\mathbb{C}^d$ each.
A translationally-invariant (TI) $k-$body Hamiltonian is defined as the operator
\begin{align}
    H = \sum_{i=1}^{N}h_i,
\end{align}
where each term $h_i$ is the same Hermitian operator, and simply translated such that the support of $h_i$ begins at site $i$.
Further, we consider periodic boundary condition in $H$, which means that the site indices are taken modulo $N$. 
Additionally, we assume \textit{geometric} locality in the Hamiltonian, which ensures that the support of $h_i$ is on $k$ sites in proximity of site $i$.
For instance, $h_1$ has support on the 1$^{\text{st}}$ to the $k^{\text{th}}$ site, $h_2$ in the 2$^{\text{nd}}$ to the $(k+1)^{\text{th}}$ site, and so on.
Whenever we use the word \emph{local} we  always mean \emph{geometrically local} (distinguished from what is sometimes called ``$k$-locality'', where  every $k$ particles partake in the interaction irrespective of their positions in space).
In this work, we will assume the system to be frustration-free, which means that the ground state of $H$ also minimizes the energy  of every local term $h_i$. 

Without loss of generality, we will shift the ground state energy to be $0$, which, combined with frustration-freeness, implies both $H \succeq 0$ and $h_i \succeq 0$,
where $\succeq$ is used to denote that the matrix is positive semidefinite.

The spectral  gap is then the lowest non-zero eigenvalue of $H$ and we denote it by  $\Delta_N$. 
Of particular relevance is the behavior 
of $\Delta_N$ as the system size increases. 
A model is defined to be \textit{gapped} if the spectral gap does not close in the thermodynamic limit, that is $\Delta_{N\to\infty} = \Delta > 0$.

We now make use of a well-known connection between the existence of a spectral gap and a quadratic operator inequality in terms of $H$: 
A frustration-free Hamiltonian has a gap greater or equal to $\delta$ if and only if the Hamiltonian satisfies the condition
\begin{align}\label{eq:gap_condition}
    H^2-\delta H\succeq 0.
\end{align} 
The above condition enforces that $H$ has no eigenvalues in the interval $(0,\delta)$.
Let us stress that all the most commonly-used methods to prove the gap of a frustration free system rely on establishing condition~\eqref{eq:gap_condition} for a given value of $\delta$~\cite{knabe_energy_1988,anshu_improved_2020,nachtergaele_spectral_1996,kastoryano_divide_2018,lemm_existence_2020,lemm_aklt_2019,pomata_demonstrating_2020}.

The condition in \cref{eq:gap_condition} allows us to formulate the gap estimation problem as an optimization problem, which turns out to be a semidefinite program (SDP).
\begin{align}\label{eq:exactPrimal}
\Delta_N & =  \max  \delta \, \, \nonumber \\
\text{s.t.} &  \quad
H^2 - \delta H \succeq 0,
\end{align}
under the condition that the ground state energy of $H$ is zero.
The optimal value of this SDP is exactly the energy of the first excited state of $H$.
Additionally, at the optimal solution, both the ground state and the first excited state space of $H$ lie in the ground state space of $H^2-\Delta_N H$.

In the general case, solving this SDP is as hard as exactly diagonalizing the Hamiltonian, since this problem imposes a positivity constraint on a matrix of size $(d^N\times d^N)$.
As $N$ increases, very quickly it becomes computationally intractable to solve. 
In the upcoming section, we show how to relax the problem of computing exact spectral gaps.
The resulting  relaxation is a \textit{local}, $n$-body, problem the solution to which is a lower bound on  the gap of $N$ site periodic system for all $N \geq 2n$, and therefore holds for the infinite system as well. 

\subsection{Simplifications of the general SDP}\label{sec:relaxations}
\begin{figure}
    \centering
    \includegraphics[trim={0 0 0 0},clip, width=0.45\textwidth]{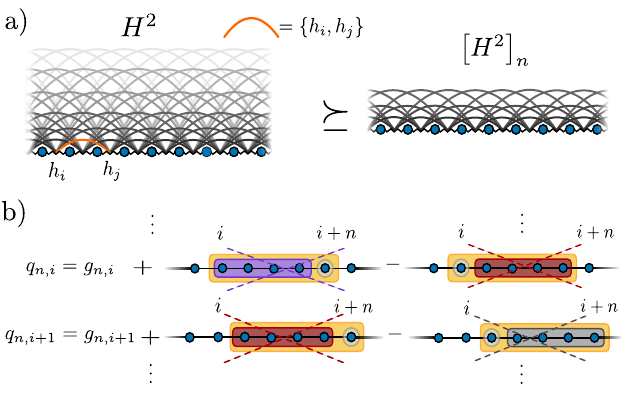}
    \caption{
        Illustration of the two relaxation steps.
        (a) Geometrical locality condition: The terms discarded from $H^2$ in \cref{eq:H_2_n} are  positive semi-definite since $h_i$ and $h_j$ have disjoint support.
        (b) Freedom in choosing a local generating term: Two local terms $g_n$ and $q_n$ generating the same global operator can differ as in \cref{eq:Y_condition} as this difference cancels telescopically.
        Inside the $n=5$-site support of each local term (orange rectangle), the operators $Y_{n-1}$ are depicted as rectangles of different colors and the identity operator is depicted by a gray circle.
         When summed over all  $i$,  the terms colored identically  cancel out as indicated by the  crosses. 
         The purple and gray terms are canceled by the preceding   and the following terms in the sum respectively.
    }   
    \label{fig:relaxation_steps}
\end{figure}

In this subsection we explain how the gap problem in an infinite translationally invariant system can be solved locally.
To do this  we will first truncate the operator in~\cref{eq:gap_condition}, replacing it with an $n$-local operator 
by omitting some positive terms from $H^2$. 
The positivity of the omitted terms will ensure that we still obtain a lower bound on the gap in the end.
Second, we will formulate a finite size ($n$-body) problem which looks for a positive  local term   which, when  summed over all translations, gives rise to the truncated $H^2 - \delta H$ operator.
Let us explain those two steps in detail (cf. \cref{fig:relaxation_steps}). 

\textbf{Geometric Locality condition.} 
The first problem we need to address is that $H^2$ is not local: it consists of products of interactions terms $h_i,h_j$ arbitrarily far apart. 
We therefore introduce a truncated operator denoted by $\left[H^2\right]_n$ which is $n$-local,   by removing terms that act on spins farther than  $n$ sites apart, namely
\begin{align}
\label{eq:H_2_n}
\begin{split}
\left[H^2\right]_n & =   \sum_{|i-j| < n-k} h_i h_j
\\
    &=\sum_i\left(h_i^2+\sum_{i<j<i+n-k}\{h_i,h_j\}\right)
\end{split}
\end{align}
(recall that $k$ is the locality range of the interaction term).
If $n$ is chosen suitably (i.e. $n \geq 2k+1$),  terms $h_i$ and $h_j$  with $j\geq i+n$ act on disjoint sets of spins and thus every  discarded term is positive semidefinite 
(as a  tensor product of positive terms).
It follows that, if we define the following operator,
\begin{align}
    Q_n(\delta) &:= \left[H^2\right]_n-\delta H \,
\label{eq:Q_n}
\end{align}
the positivity of $Q_n (\delta)$   implies positivity of $H^2 - \delta H$. 
Hence, we can prove  condition~\eqref{eq:gap_condition} by showing $Q_n(\delta)\succeq 0$. 

\textbf{Positive semidefinite generating local term.} 
Next, we will try  to prove $Q_n(\delta)\succeq 0$ by looking  for an $n$-body local term $q_n$ which is positive semidefinite such that $Q_n(\delta)$ can be written as 
\begin{align} 
\label{eq:q_n}
Q_n (\delta) = \sum_i q_{n,i} \, \, , \quad 
q_{n,i} \succeq 0  
\end{align}
where the the $i$ subscript indicates the position of the first spin on which $q_n$ acts (the $n$ subscript simply denotes the  $n$-locality, i.e.\ that $q_n$ is supported on $n$ neighboring sites).
If for some value of $\delta$, $Q_n(\delta)$ is positive semidefinite,  we can hope to find such a positive generating term.
It could, however, be the case that a local positive generating term $q_n$ can only be found for a lower value of $\delta$, and thus we are further relaxing the problem in this step. (In practice our numerical demonstrations below show that the effect of this relaxation step (\cref{eq:q_n}) is negligible.) 

Note that if~\eqref{eq:q_n} holds for a given support $n$ and a global system size $N$, then the same decomposition will hold for any $N' > N$. 
Hence, finding such a decomposition with a positive $q_n$ serves as a proof for the infinite system as well. 

We will now proceed to show how the condition~\eqref{eq:q_n} 
can be expressed in terms of local operators, and thus arrive at a finite size relaxation of~\cref{eq:exactPrimal} that is numerically tractable for any finite support $n$.
\subsection{Primal SDP}
\label{sec:primal_sdp_formulation}
After substituting the operator $H^2$ in the exact SDP \cref{eq:exactPrimal} with $[H^2]_n$, and imposing the condition that $Q_n(\delta)$ has a positive operator decomposition we arrive at the following problem
\begin{align}
  & \max_{\delta, q_n}  \delta \, \, \nonumber \\
\text{s.t.} &  \quad
Q_n (\delta) = \sum_{i = 1}^N q_{n,i} ,  \label{eq:global recovery} \\
& \quad q_{n} \succeq 0  \,.
\end{align}
At this stage, the constraint \cref{eq:global recovery}   still involves the exponentially-big operator $Q_n (\delta)$. 
We now argue that the condition~\eqref{eq:global recovery} is equivalent to a local one, involving only operators with support on $n$ sites. 
First, notice that $Q_n (\delta)$ can be written in terms of a    local generating term $g_n$ as  $Q_n (\delta) = \sum_i g_{n,i} (\delta)$  
where 
\begin{equation}
\label{eq:g_n}
g_{n} (\delta) :=\left(h_1^2+\sum\limits_{1<j<1+n-k}\{h_1,h_j\}-\delta h_1\right)   \, . 
\end{equation}
The above term  $g_{n}$ will in general  not be positive semidefinite, even for a globally positive semidefinite operator $Q_n (\delta)$. 
Moreover the choice of local generating term is not unique as we can choose different ways to distribute  the coefficients of the terms in $g_n$ as long as they sum up to the right amount in the global operator. 
For instance, the two local terms, $h_i = Z_iZ_{i+1}+X_i$ and $h'_i = Z_iZ_{i+1}+\frac{1}{2}X_i+\frac{1}{2}X_{i+1}$,
 both give rise to the transverse field Ising model Hamiltonian when summed over all sites in the system.
This freedom of choice can be completely characterised, 
which is done in Appendix~\ref{app:LTI_proof}. 
In particular, we show that two generating local terms $q_{n}$ and $g_{n}$ sum up to the same global operator, if and only if they are related by
\begin{equation}\label{eq:Y_condition}
q_{n} =g_{n} +\mathbbm{1}\otimes Y_{n-1} - Y_{n-1}\otimes\mathbbm{1}    \, ,
\end{equation}
where $Y_{n-1}$ can be any operator supported on $n-1$ sites. 
Condition~\eqref{eq:Y_condition} is chosen such that, when summed over $i$, the terms involving $Y_{n-1}$ cancel telescopically. 
It follows that the constraint~\eqref{eq:global recovery} corresponds to looking for a suitable $Y_{n-1}$ that makes the local generating term $q_{n,i}$ positive semidefinite. 
In this way, we arrive at the final form of our SDP, which reads
\begin{align}
\label{eq:primal_TI_Y}
\delta_{\mathrm{LTI}}(n) &= \max_{\delta,Y_{n-1}} \; \delta \\ \label{eq:primal_TI_constraint}
    \text{s.t. } & \quad 
    g_n(\delta) + 
    \mathbbm{1}\otimes Y_{n-1} - 
    Y_{n-1}\otimes\mathbbm{1}
    \succeq 0, 
\end{align}
where $g_n$ can be any local generating term for $Q_n(\delta)$, for example the one in \cref{eq:g_n}. Any  two generating terms are related by some $Y_{n-1}$ terms as in \cref{eq:Y_condition}, and we anyway  optimize over all possible choices of those terms.

We will refer to the problem we arrived at in \cref{eq:primal_TI_Y} as the locally translationally invariant (LTI) gap SDP. It is a semidefinite program which involves a positivity constraint on an $n$-body (i.e.\ $d^n\times d^n$ dimensional) matrix. 
To obtain a positive decomposition of $Q_n(\delta)$, one option is to formulate it as a sum-of-squares (SOS) problem~\cite{lasserre_convexity_2008,parrilo_semidefinite_2003,parekh_application_2021,barak_hypercontractivity_2012,navascues_convergent_2008}, where positivity is enforced through moment matrices associated with a chosen set of monomials. In contrast, we directly optimize the matrix $q_n$ itself, making use of an exact characterization of the degrees of freedom that leave the sum $Q_n(\delta)$ unchanged.

The LTI SDPs form a \textit{hierarchy} of optimization problems, and the level of the hierarchy is labeled by the size $n$ of the finite system.
As the size $n$ of the local operators increases, the solution $\delta_{\mathrm{LTI}}(n)$ is guaranteed not to decrease, forming a sequence of lower bounds on the gap of the infinite system
\begin{align}
    \delta_{\mathrm{LTI}}(n)
    \leq
    \delta_{\mathrm{LTI}}(n+1)
    \leq \ldots \leq 
    \Delta_{N\to\infty} \, .
\end{align}
This follows from the fact that the operator $q_{n+1}$ supported on $n+1$ sites embeds the solution $q_n$ obtained using $n$ sites, and they are related as
\begin{align}
    q_{n+1} = a \cdot \mathbbm{1} \otimes q_n + b \cdot q_n \otimes \mathbbm{1} + c \cdot S,
\end{align}
where $ a, b, c \geq 0$  and $S\succeq 0$.
Thus, the SDP provides more accurate lower bounds as $n$ increases, albeit at a higher computational cost.

\subsection{Dual SDP}
\label{sec:dual_sdp_formulation}
Any semidefinite program which is defined as a maximization has a dual formulation in terms of a minimization problem~\cite{boyd_convex_2023}.
Importantly, the optimum of the dual serves as an upper bound on the primal optimal. 
When the two optima coincide, the problem exhibits so-called strong duality and one can equivalently solve either one of the problems. 
In this subsection we present the duals of the  spectral gap SDPs introduced in the previous subsections and discuss   several useful insights which become transparent when  the problem is viewed from the  perspective of the dual.

The dual program of the exact spectral gap SDP of~\cref{eq:exactPrimal} is 
\begin{align}
\label{eq:exactDual}
    \Delta_N = \min_{\rho} &\;\Tr(H^2\rho) \\ \nonumber
    \text{s.t. } & \rho \succeq 0 \\ \nonumber
    & \Tr(H \rho ) = 1 .
\end{align}
Note that this problem attains the optimum of the primal~\cref{eq:exactPrimal}, i.e.\ the exact gap of the $N$-body system $\Delta_N$:
The minimizer is $\rho = \ket{\psi_1}\bra{\psi_1}/\Delta_N$, where $\ket{\psi_1}$ denotes the first excited state of $H$.

In view of \cref{eq:exactDual}, we can now justify   the first relaxation step \cref{eq:H_2_n}, in which we truncated $H^2$ by omitting the  terms $h_i\otimes h_j$ for $|i-j|>n$. 
We argue that the contribution of such highly nonlocal terms to the objective in \cref{eq:exactDual} can be, in some cases, negligible. 
This is because of the local nature of excitations in gapped systems. 
As   shown in Ref.\ \cite{haegeman_elementary_2013},  an excited state ${\psi_{p,\alpha}}$  with  momentum $p$ and energy $E_{\alpha}$,  which is separated by an energy gap from above and from below, can be well approximated by a momentum superposition of an $l$-local operator $O^{(l)}$  acting on     the ground state $\psi_0$:
\begin{equation}
\label{eq:approx_momentum_ket}
    \ket{\psi_{p,\alpha}} \approx \sum_x e^{ipx} \tau_x(O^{(l)}) \ket{\psi_0} , 
\end{equation}
where $\tau_x$ is the translation operator by $x$ lattice sites, 
 and where the approximation error vanishes exponentially with $l$ (\cite{haegeman_elementary_2013}, theorem 1). 
We thus see that for a frustration free Hamiltonian, the expectation  value of  a term $h_i\otimes h_j$ for $|i-j|>l$ in the state on the RHS of \cref{eq:approx_momentum_ket} is identically zero  because   in every term in the sum either $h_i$ or $h_j$ acts on the ground state $\psi_0$ and annihilates it.
The  result of this analysis is that for gapped frustration-free Hamiltonians, which in addition have an energy gap between the first  and second excited states,
the overall error in the  expectation value   $\bra{\psi_{p,\alpha}}H^2\ket{\psi_{p,\alpha}}$ when $H^2$ is replaced with its truncated version  $[H^2]_n$ decays exponentially with the distance $n$.
While this does not guarantee that replacing $H^2$ with $[H^2]_n$ in \cref{eq:exactDual} will produce a result close to the exact gap (as the minimizer there can be different from $\ket{\psi_{p,\alpha}}$), it certainly provides an intuition for why this relaxation step is reasonable, despite the fact that when truncating we disregard a vast amount of terms in $H^2$. In practice we observe that the reduction in the detected gap due to this relaxation is indeed minimal (see \cref{fig:AKLT} below).


The dual of the LTI  gap SDP \cref{eq:primal_TI_Y} has the following form
\begin{align}
\label{eq:dual_TI_Y}
    \delta_{\mathrm{LTI}}(n) = \min_{\rho_n} &\; \Tr \left[\rho_n \left(h_1^2 + \sum_{j=2}^{n-1}\{h_1,h_j\} \right)\right] \\ \nonumber
    \text{s.t. } & \rho_n\succeq 0 \\ \nonumber
    &  \Tr_1(\rho_n) = \Tr_n(\rho_n)  \\
    & \Tr(\rho_n h_1) =1,\nonumber
\end{align}
where the optimization variable $\rho_n$ is the reduced state  on $n$ consecutive sites.
The constraint
\begin{align}
    \Tr_1(\rho_n) = \Tr_n(\rho_n)
\end{align}
indeed requires $\rho_n$ to obey local translation invariance and it arises as a dual to the constraint \cref{eq:primal_TI_constraint}. 


Due to strong duality, the following relation is satisfied between the primal and dual optimal variables in \cref{eq:primal_TI_Y,eq:dual_TI_Y} respectively
\begin{align}
\label{eq:compl_slack}
    \rho_n q_n = 0.
\end{align}
This relation which is known as complementarity slackness has several useful consequences. 
Among others, it provides us with the gradient of the solution to the SDP with respect the parameters defining the problem data, without incurring further computational cost \cite{reehorst_navigator_2021}. 
In our case the problem data is the Hamiltonian interaction term $h$ and, similarly to the   Hellmann--Feynman theorem,   the gradient with respect to $h$ of the solution to  \cref{eq:dual_TI_Y}   can be computed from  the minimizing state:
$$\nabla_h \delta_{\mathrm{LTI}}(n) = \Tr(\rho_n \nabla_h (q_n)) .$$
This could be used to maximize the gap over families of parent Hamiltonians of a given ground state. We elaborate on this in \Cref{app:SDP_gradient}.

\section{Connection to previous finite-size criteria}
\label{sec:existing_methods}
In this section, we compare our method to finite-size criteria-based techniques, which are the prevalent methods used to lower bound spectral gaps.
We find that the  LTI SDP \cref{eq:primal_TI_Y} will always detect a gap greater or equal to the ones detected by existing finite size methods. 
This is simply because each finite size method looks for a specific $n$-local positive decomposition, whereas our SDP optimizes over all such decompositions. 
In this section, we prove that the decompositions corresponding to different finite size methods are feasible (yet not necessarily optimal) in our SDP. 
In other words, they are included in our search.
Additionally, the finite-size method requires the interaction terms in the Hamiltonian to be projectors, whereas the LTI SDP works directly with the original interactions. 
While using the projector form does not affect the condition for the existence of the gap, it reduces the accuracy of the resulting bound.

Note that the lower bounds based on finite size criteria pertain to the spectral gaps $\Delta_N$ of systems with periodic boundary. 
Their expressions, however, involve the computation of spectral gaps for open boundary Hamiltonians, which we denote as $\epsilon_n$.
For consistency with the formulation of the existing methods~\cite{knabe_energy_1988,gosset_local_2016}, we restrict the discussion to nearest-neighbor projector Hamiltonians
\begin{align} \label{eq:projectors}
    H = \sum_ih_i, \quad h_i^2=h_i.
\end{align}
Importantly, if the local term $h$ of a given frustration-free Hamiltonian is not a projector, it can always be modified into a projector Hamiltonian by adjusting the eigenvalues of the subspace orthogonal to the ground space of $h$. 
It follows from a simple argument that proving the gap of this second Hamiltonian implies a gap for the first as well~\cite{gosset_local_2016}(c.f. \Cref{sec:numerics}).
We will now prove that the set of  feasible local terms in the  LTI-SDP \cref{eq:primal_TI_Y} includes the ones obtained from   known  finite size criteria.
Thus, the value of  LTI SDP achieves a tighter lower bound on the spectral gap.
\subsection{Finite-size criteria}
\label{sec:finite_size}
In the following discussion, we consider a nearest neighbor Hamiltonian comprised of local terms that are projectors, as in \eqref{eq:projectors}.
We review the existing methods based on finite-size conditions, classifying them into two types - based on (i) finite system gaps, and (ii) local ground state projectors.

\textbf{(Method 1) Based on finite system gaps.}
First formulated by Knabe~\cite{knabe_energy_1988}, these criteria provide a way to relate the spectral gap of a finite open-boundary system to the spectral gap in the thermodynamic limit.
The open boundary Hamiltonian on a finite system of size $n$ is,
\begin{align}
    H_n^{\mathrm{OBC}} = \sum\limits_{i=1}^{n-1}h_{i} ,
\end{align}
If $H_n^{\mathrm{OBC}}$ has a spectral gap of $\epsilon_n$~\footnote{Note that for ease of reading, we denote spectral gaps of periodic boundary Hamiltonians with $\Delta$ and of open boundary Hamiltonians with $\epsilon$.}, then the lower bound to the gap in the thermodynamic limit is given as
\begin{align}
    \delta(n)=c_n\left(\epsilon_n-t_n\right).
    \label{eq:finite_size_bound}
\end{align}
Here $c_n$ is a function of $n$ which does not affect the condition for existence of the gap. 
The quantity $t_n$ defines a threshold, i.e., if the local gap satisfies the condition $\epsilon_n>t_n$, then the system is gapped in the thermodynamic limit.
If $\Delta_m$ is the gap of periodic system on $m$ sites, then the Knabe bound~\cite{knabe_energy_1988} is
\begin{align}
\label{eq:knabe_bound}
    \Delta_m\geq\delta_{\mathrm{Knabe}}(n) = \frac{n-1}{n-2}\left(\epsilon_n-\frac{1}{n-1}\right)
\end{align}
for $m>n>2$. 
Similarly, the Gosset--Mozgunov bound~\cite{gosset_local_2016} is
\begin{align}
\label{eq:gosset_bound}
    \Delta_m\geq\delta_{\mathrm{GM}}(n)=\frac{5}{6}\frac{n^2+n}{n^2-4}\left(\epsilon_n-\frac{6}{n^2+n}\right)
\end{align}
for $m>2n>4$.
Comparing the bounds with \cref{eq:finite_size_bound}, the functions $c_n$ and $t_n$ satisfy the following properties
\begin{align}
    \lim_{n\to\infty}c_n\leq 1,
    \quad\lim_{n\to\infty}t_n=0.
\end{align}
such that
\begin{align}    \lim_{n\to\infty}\delta_{\mathrm{Knabe},\mathrm{GM}}(n)\leq\lim_{n\to\infty}\epsilon_n.
\end{align}

\textbf{(Method 2) Based on ground state projectors.}

This method was introduced by Nachtergaele, Fannes, and Werner in Ref.~\cite{nachtergaele_spectral_1996,fannes_finitely_1992}, and is also known as the ``martingale'' criterion.
Instead of using the gap of a finite size system, the presence of a gap is established by checking if a relation in terms of the ground space projectors is satisfied.
The bound is obtained by first constructing a new coarse-grained Hamiltonian $\tilde{H}=\sum_{\mathcal{I}}\tilde{h}_{\mathcal{I},\mathcal{I}+1}$ obtained by blocking $l$ sites into a single site, i.e.
\begin{align}
\label{eq:coarse_grain}
    &\mathcal{I}=\{(i-1)l+1,...,il\}\\ \nonumber
    &\tilde{h}_{\mathcal{I},\mathcal{I}+1} = \sum\limits_{j=(\mathcal{I}-1)l+1}^{(\mathcal{I}+1)l-1}h_{j,j+1}.
\end{align}
The locality of the new Hamiltonian $\tilde{h}_{\mathcal{I},\mathcal{I}+1}$ is $2l$-local in the original system.
Using \cref{eq:coarse_grain}, we have the following relation between the original and the blocked Hamiltonian, as in the latter every term form the former appears either once or twice:
\begin{align}
& H\succeq\frac{\tilde{H}}{2}.
\end{align}
W.l.o.g, the coarse-grained Hamiltonian $\tilde{h}$ can be converted into the corresponding projector Hamiltonian $K = \sum_{\mathcal{I}}P_{\mathcal{I},\mathcal{I}+1}$, which satisfies the relation,
\begin{align}
\tilde{h}_{\mathcal{I},\mathcal{I}+1}\succeq\epsilon_{2l}P_{\mathcal{I},\mathcal{I}+1},
\end{align}
where $\epsilon_{2l}$ denotes the gap of $\tilde{h}$.
To compute the martingale bound,
we first state the inequality from Lemma 6.3(2) of~\cite{fannes_finitely_1992},
\begin{align}
    \{P_{\mathcal{I},\mathcal{I}+1},P_{\mathcal{I}+1,\mathcal{I}+2}\}\succeq -\eta_l(P_{\mathcal{I},\mathcal{I}+1}+P_{\mathcal{I}+1,\mathcal{I}+2}),
\end{align}
where
\begin{align}
\label{proj_norm_to_bd}
    \eta_l = \vert\vert(\id-G_{1,\ldots,2l})  (\id-G_{l+1,\ldots,3l})  - (\id-G_{1,\ldots,3l})  \vert\vert,
\end{align}
and $G_{i,i+1,...,i+m-1}$ denotes the projector onto the ground space of the $m$-site Hamiltonian with open boundary conditions
\begin{align}
    H_{m,i}^{\mathrm{OBC}}=\sum_{j=i}^{i+m-1}h_j.
\end{align}
Finally, using the above inequality, the bound on the gap of the original Hamiltonian is given as,
\begin{align}
\label{eq:martingale_bound}
    \delta_{\mathrm{M}}(3l) = \frac{\epsilon_{2l}}{2}(1-2\eta_l).
\end{align}
The lower bound on gap of $K$ is $(1-2\eta_l)$ which is then rescaled by $\epsilon_{2l}/2$ to recover the bound for the original Hamiltonian $H$.
Therefore, to prove the existence of a gap it is sufficient to find a value for $l$ that satisfies the condition $\eta_l<0.5$, which is useful in cases where analytical proofs are required~\cite{brandao_local_2016,haferkamp_improved_2021}.
Given the form for the ground state projectors, it is possible to compute $\eta_l$ numerically by increasing the value of $l$ until the threshold of $\eta_l<0.5$ is achieved.
The other possibility is to analytically bound $\eta_l$ as
 $\eta_l\leq\eta'(l)$, where the function $\eta'(l)$ becomes small enough for large $l$.
 This is problematic if the entire gap estimate (i.e. $\delta_{\mathrm{M}}$) is required, because if $\eta'(l)<0.5$ is only reached for a large value of $l$,  then it becomes harder to compute the prefactor $\epsilon_{2l}$ in \cref{eq:martingale_bound}, as it requires the exact diagonalization of a $2l$-sites system. 

\subsection{Proof of inclusion}
We will now prove that the LTI SDP provides more accurate and larger estimates for the thermodynamic limit gap than the finite size criteria bounds.
More precisely, we obtain the relation,
\begin{align}
\label{eq:gap_relation_LTI}
    \delta_{\mathrm{LTI}}(n)\geq\delta_{\mathrm{Knabe,GM,M}}(n),
\end{align}
meaning that the Knabe, Gosset--Mozgunov, and "martingale" bounds (\cref{eq:finite_size_bound,eq:martingale_bound}) calculated for a particular size $n$ of the truncation, yield weaker estimates, than the LTI SDP of the same size. 
Concretely, we show that the finite size criteria bounds lie in the feasible set of the LTI SDP.
The above relation relies on finding a positive semidefinite generating term $q_n(\delta)$ for the truncated operators $Q_n(\delta)$ corresponding to each of the bounds.
We show the inclusion for the martingale bound using the following positive local decomposition,
\begin{align}
\begin{split}
    q_{3l}(1-2\eta_l):=\eta_l(P_{\mathcal{I},\mathcal{I}+1}+P_{\mathcal{I}+1,\mathcal{I}+2})
\\+\{P_{\mathcal{I},\mathcal{I}+1},P_{\mathcal{I}+1,\mathcal{I}+2}\},
\end{split}
\end{align}
which assumes evaluating the LTI SDP for the coarse-grained Hamiltonian $K$ (as defined in Method 2 (\cref{sec:finite_size})).
Similarly, the expression for the Knabe bound is,
\begin{align}
    q_n(\delta_{\mathrm{Knabe}}(n)) &:= \left(\frac{1-\epsilon_n}{n-2}\right)h_{n,i} \nonumber \\
    & + \sum_{x=1}^{n-2} \frac{1}{n-x-1} \sum\limits_{j=0}^{n-x-2}\{h_{i+j},h_{i+j+x}\}.
\end{align}
The detailed proofs of these claims as well as of the attainment of the Gosset--Mozgunov bound are given in Appendix~\ref{app:special_case}.

An important implication of \cref{eq:gap_relation_LTI} is that the LTI-SDP does not necessarily  lower bound the gap of the open boundary system, i.e. $\epsilon_n$, in the thermodynamic limit.
This is because the formulation of $\delta_{\mathrm{LTI}}(n)$ only ensures that it lower bounds the gap of a periodic ring of size $m>2n$.
This becomes evident by taking the $n\to\infty$ limit on both sides of \cref{eq:gap_relation_LTI}, 
\begin{align}
    \lim_{n\to\infty}\delta_{\mathrm{LTI}}(n)
    \geq
    \lim_{n\to\infty}\delta_{\mathrm{Knabe}}(n) = \lim_{n\to\infty}\epsilon_n,
\end{align}
shows that $\delta_{\mathrm{LTI}}(n)$ upper bounds $\epsilon_n$ in the limit. 
This is in agreement with the well known fact that gappedness of OBC system implies the gappedness of PBC system, while the converse is not always true.
An open question persists as to whether $\delta_{\mathrm{LTI}}$ converges to the gap in the thermodynamic limit.
\section{Numerical benchmarking}
\label{sec:numerics}
In this section we present the results of the numerical application of the method presented in \Cref{sec:method} to 1-dimensional spin-chain models. 
We compared our method with existing  finite-size criteria, namely, the   \mbox{Knabe} bound (\cref{eq:knabe_bound})  and the bound due to Gosset and Mozgunov (\cref{eq:gosset_bound}). 

Because of the inclusions shown in Sec. \ref{sec:existing_methods}, we know that for the same $n$ our method always gives an equal or tighter lower bound on the gap than either of those existing bounds.
However, when comparing the methods we should keep in mind that  the computational tasks involved are different.
While the finite-size methods require the computation of the  gap of the $n$-body OBC Hamiltonian---a task with a memory scaling of  $\mathcal{O}(d^n)$---our method  optimizes over a positive semidefinite matrix of size $d^n\times d^n$. 
The results in this section show that our method outperforms the existing ones even when compared on equal footing (in terms of size of the largest variable in the problem):
For the models on which we benchmarked, we observed that $\delta_{\mathrm{LTI}}(n) \geq \delta_{\mathrm{Knabe}}(2n)$, and similarly for $\delta_{\mathrm{GM}}(2n)$. 

Another difference between our method and the existing ones is that the finite size criteria assume that the interaction term in the Hamiltonian is a projection.  
This assumption does not restrict the generality of the methods for the task of \emph{detecting} a gap, since one can always bound the gap of the original Hamiltonian $ \sum_i h_i$ by the gap of the interaction term $h_i$ times the gap of of the projectors Hamiltonian $ \sum_i \Pi_i$, where $\Pi_i$ is the projector orthogonal to the ground space of $h_i$. However, when the optimization is restricted to be over the new projector Hamiltonian,
some information about the exact spectrum of the Hamiltonian is lost in the estimation.
Favorably, our method does not need the assumption that local terms have to be projectors. 
In our numerical benchmarking, we observed situations where better bounds were obtained by solving the LTI SDP with the original Hamiltonian term than with the projector term.
 

Before presenting the results we address a technical but important issue: 
When performing  the optimization on a computer the SDP solver will produce a solution which is feasible up to some finite precision. 
If we wish to regard such a solution as providing a mathematical proof of the existence of a gap, we have to  make sure that we have a way to obtain a strictly feasible solution from the one produced by the solver.
In our case we obtain a numerical solution for the variable $q_n$ which should serve as our certificate of a gap (via equation   \cref{eq:global recovery}). 
In practice we obtain $q_n \succeq -\epsilon $ where $\epsilon>0$ is the solver precision. 
For the LTI SDP one cannot simply add $\epsilon \id$ to $q_n$ as this would shift the global energy by an extensive amount and a more refined approach has to be taken. 
In \Cref{app:certify_solution}, we explain how we modified the  LTI SDP to  obtain   certified solutions.
When testing the method on gapless frustration free models and ones with very small gaps, we observed that the correction detailed in \Cref{app:certify_solution} is necessary.  Without it the solver may produce wrong results (i.e.\ detect a finite gap when the model is gapless or detect a gap larger than the true one).

\subsection{The AKLT model}
The AKLT model was the first example of a provably gapped isotropic spin chain with integral spin~\cite{affleck_rigorous_1987}.
When the model was initially introduced it was proven to be gapped but an explicit lower bound was not provided~\cite{affleck_valence_1988}.
In the same year, Knabe introuced his method and provided a lower bound of $0.24806$ for the gap of the AKLT model. 
Garcia-Saez et al.~\cite{garcia-saez_spectral_2013}, using a method based on the rotation symmetry of the Hamiltonian,  estimated the AKLT gap to be $0.350$,
which is compatible with both exact diagonalization results for finite periodic chains
$(0.35012$ for $n=16$~\cite{garcia-saez_spectral_2013}), 
as well as DMRG calculations for finite open chains ($0.35037$ for $n=100$ \footnote{Mingru Yang, private communication}). 
Our method proves a lower bound of $\Delta_{\text{AKLT}} \geq 0.34976$ 
which was obtained by solving  the LTI SDP \cref{eq:primal_TI_Y} for $n=6$. 
\begin{figure}[!b]
    \centering
    \includegraphics[trim={0 0 0 0},clip, width=0.45\textwidth]{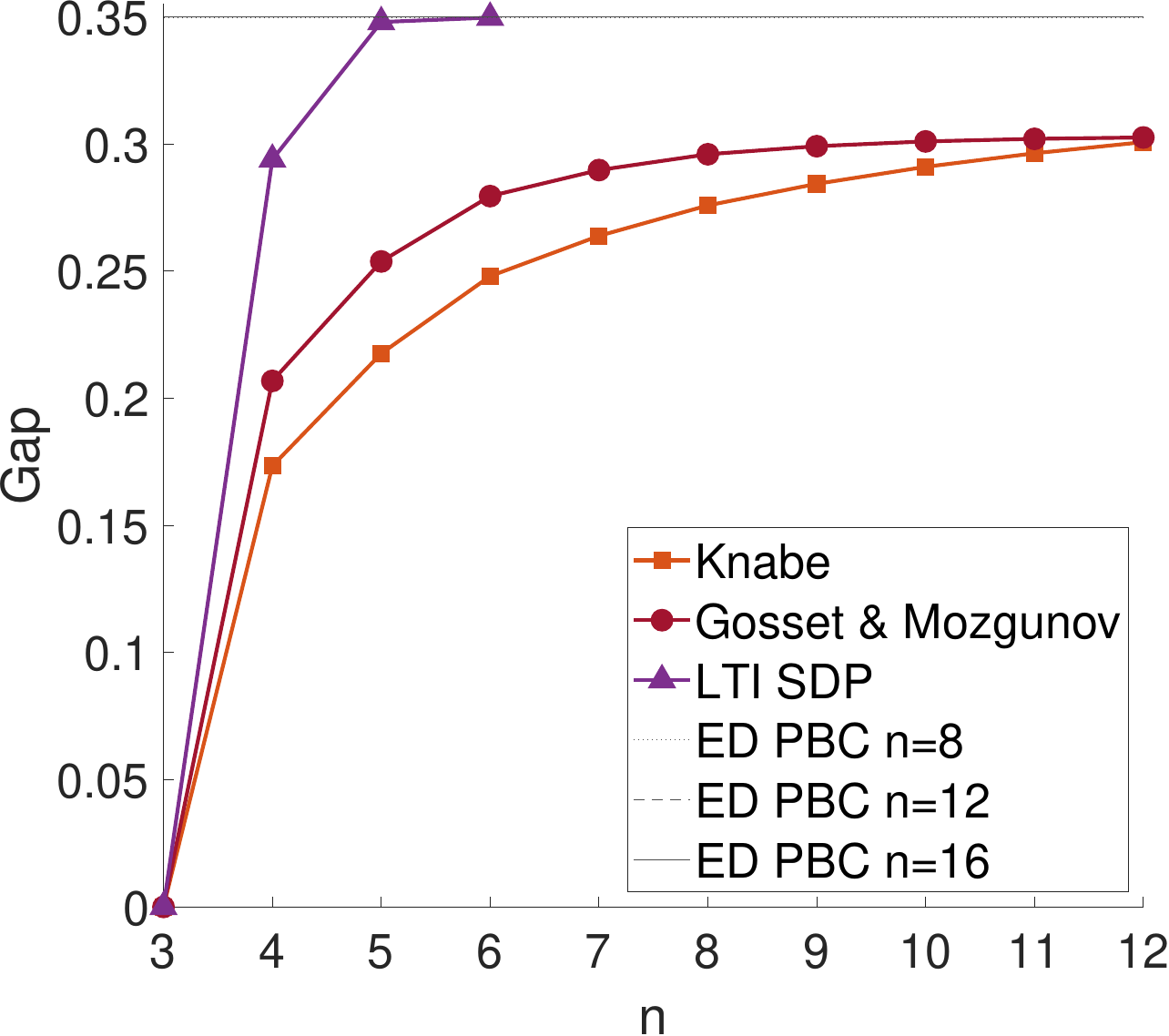}
    \caption{The gap of the $S=1$ AKLT chain.
    Lower bounds on the energy gap above the ground state of the infinite AKLT chain obtained with    different methods are plotted as a function of the size $n$ of the finite system: The Knabe bound \cref{eq:knabe_bound}, the  Gosset--Mozgunov bound \cref{eq:gosset_bound}, and our LTI SDP method \cref{eq:primal_TI_Y}.
    In addition, gaps of finite systems with   periodic boundary conditions (PBC) obtained with exact diagonalization (ED) are plotted for rings of sizes $n=8,12$, and $16$ by dotted, dashed, and solid horizontal black   lines. Note that those lines lie essentially on top of each other in the plot (the value for $n=12$ and $n=16$ differs in the 6th digit).
    The LTI SDP significantly outperforms the other two methods and gives a lower bound of $0.34976$ for the AKLT gap.
    }
    \label{fig:AKLT}
\end{figure}

\Cref{fig:AKLT} shows the Knabe (\cref{eq:knabe_bound}) and Gosset--Mozgunov (\cref{eq:gosset_bound}) lower bounds, 
and the lower bounds computed with our method (the solution of \cref{eq:primal_TI_Y}), all for different values of $n$.
In addition the finite system gaps obtained using  exact diagonalization  for  periodic boundary conditions   (PBC) are plotted for $n=8,12$ and $16$ as horizontal lines. 
Note that the exact diagonalization  results for the PBC chain are neither  lower nor upper  bounds on the gap in the thermodynamic limit. 
They are plotted for reference because the Knabe and Gosset--Mozgunov bounds, as well as the LTI SDP all provide lower bounds to the finite periodic systems' gaps as well.

We observe that while all three methods (ours, Knabe, and Gosset--Mozgunov) detect the gap starting from  system size $n=4$, the LTI SDP performs significantly better than any of the other methods. 
This is true even if we compare $\delta_{\mathrm{LTI}}(n)$ with $\delta_{\mathrm{Knabe}}(2n)$ and $\delta_{\mathrm{GM}}(2n)$. 
Recall that the LTI SDP \cref{eq:primal_TI_Y}  provides a lower bound on the gap of any periodic system of size $\geq 2n$. 
Our results for the AKLT model show that the LTI SDP gets very close to this optimal value: for $n=6$ we have $\Delta_{\text{PBC}}(12) - \delta_{\text{LTI}}(6) = 0.00036$. 
\subsection{Deformed clock models}
Next we present the results of the application of our method to a family of models generated using a  generalization of Witten’s conjugation~\cite{wouters_interrelations_2021}.
In this approach one starts out with a given frustration-free Hamiltonian and deforms it  by acting on the local terms with a tensor-product deformation operator.
By applying a parameter-dependent deformation one arrives at a family of frustration-free models.

We studied the family of models arising from a deformation of the classical 3-level Potts clock model:
\begin{equation}
\label{eq:potts_model}
    h_{i,i+1} = 2-\sigma_i \sigma^\dagger_{i+1} - \sigma^\dagger_i \sigma_{i+1} , 
\end{equation}
where $\sigma_i$ is an operator that rotates the clock at site $i$ by a third of a turn. Precisely, it is determined by the $\mathbb{Z}_3$ clock algebra:
$\sigma^3 = \tau^3 =\id$, 
$\sigma^{2} = \sigma^\dagger$, 
$\tau^{2} = \tau^\dagger$, and
$\sigma\tau = \omega \tau\sigma$, where $\omega = e^{2\pi i /3}$.
This model has a gap above its    degenerate   ground space, spanned by states consisting of  tensor products  of the eigenstates of $\sigma$ for each of its eigenvalues $\omega^k,\; k=0,1,2$.

The applied deformation depends on two real parameters $(r,s)$ and the resulting family of models takes the form:
\begin{align}
\label{eq:deformed_potts}
    H(r,s) = & -\sum_i 
        \sigma_i \sigma^\dagger_{i+1} +\frac{f(r,s)}{2}(\tau_{j} + \tau_{j+1}) + \\ \nonumber
        & g_1(r,s)\tau_{j} \tau_{j+1} + g_2(r,s)\tau_{j} \tau^\dagger_{j+1} +
        \text{h.c.}
      + \epsilon ,
\end{align}
where the coefficients $f,g_1$, and $g_2$ are as follows:
\begin{align}
    f = 
    -\frac{2}{9}\left[ 
    2 \left( rs + \omega\frac{r}{s^2} + \overline{\omega}\frac{s}{r^2} \right) - 
    \left( \frac{1}{rs} + \overline{\omega}\frac{r^2}{s} + \omega\frac{s^2}{r} \right)
    \right] \\
    g_1 = 
    -\frac{2}{9}\left[ 
    \omega \left( \frac{r^2}{s} + \frac{s}{r^2} \right) + 
    \overline{\omega}  \left( \frac{s^2}{r} + \frac{r}{s^2} \right) + 
    rs + \frac{1}{rs} 
    \right] \\
    g_2 = 
    \frac{1}{9}\left[ 
    3 +  \left(  \frac{1}{rs} + \frac{s^2}{r} + \frac{r^2}{s} \right) - 
    2 \left(  {rs} +  \frac{s}{r^2} +  \frac{r}{s^2} \right)
    \right], 
\end{align}
and where $\epsilon$ is chosen such that the interaction terms are positive semidefinite.
  
Due to the fact that the deformation preserves the product-state structure of the ground states of \cref{eq:potts_model},  the martingale method can be used to prove that the model is gapped in the whole parameter range \cite{nachtergaele_spectral_1996}: the overlaps between the the ground-state projectors  are easy to  compute analytically and $\eta$ in \cref{proj_norm_to_bd} is thus shown to decay with the size of the support of the state ($l$ in \cref{proj_norm_to_bd})  (see also \cite{rozman_bounding_2024} for a specific treatment of    this family of models). 
This strategy for proving the existence of a gap, however, has the disadvantage that it does not provide us with a concrete lower bound. 
The exact lower bound proven in this way has a prefactor: the gap of the blocked interaction  term $\tilde{h}$ which acts on $2l\gg 1$ spins. 

Using Knabe's criterion, the authors of Ref.~\cite{wouters_interrelations_2021} computed lower bounds on the gaps of systems with a deformation in a region of the   $(r,s)$ parameter space (see Appendix B.2 in \cite{wouters_interrelations_2021}). 
We   benchmarked our method on this family of models to see how  it compares to the finite-size criteria in terms of the  range of parameters in which a nonzero gap is detected.  

\begin{figure}[ht]
    \centering
    \includegraphics[trim={0 0 0 15mm},clip, width=0.5\textwidth]{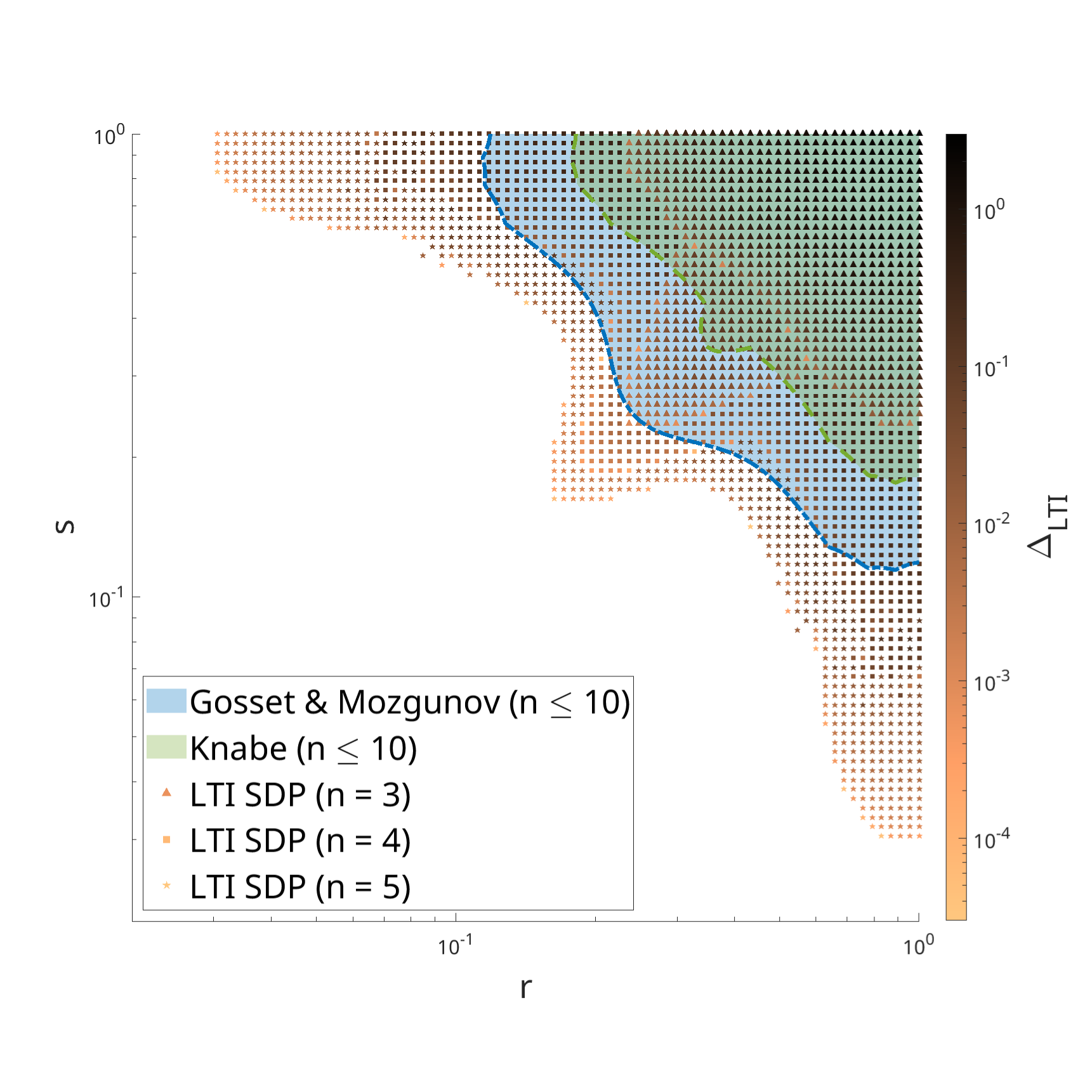}
    \caption{Gaps of the $\mathbb{Z}_3$ family of deformed Potts clock models.
    For a family of spin chain models parameterized by two parameters $r,s$ (\cref{eq:deformed_potts}) lower bounds on the gap were computed at different values of the parameters using different methods: 
    the Knabe and Gosset--Mozgunov finite size criteria (\cref{eq:knabe_bound} and \cref{eq:gosset_bound} respectively), and
    our  LTI SDP method \cref{eq:primal_TI_Y}. 
    The shaded regions indicate the parameter values for which the finite-size methods detected a gap using exact diagonalization results for systems up to size $n=10$ (green region with dashed border for Knabe and  blue region with dash-dotted border for Gosset-Mozgunov). 
    The different markers indicate parameter values for which the LTI SDP detected a gap, with different markers corresponding to different values of $n$ (see legend). 
    For each coordinate $(r,s)$, we only plot the marker corresponding to the lowest value $n$ for which the LTI SDP detected a gap.
   %
   %
    Darker marker  colors indicate larger gaps (see color bar). Each marker corresponds to the larger of the two gaps resulting from using either  the original Hamiltonian or the  modified Hamiltonian---in which the  interaction term is  a projection---as the input for the LTI SDP.
    }
    \label{fig:z3_r_s_deform}
\end{figure}

We computed lower bounds on the gap for values of $(r,s)$ in the region $(0,1]\times(0,1]$ using the Knabe method (\cref{eq:knabe_bound}), Gosset--Mozgunov (\cref{eq:gosset_bound}) and our method (\cref{eq:primal_TI_Y}).
\Cref{fig:z3_r_s_deform} shows the regions in which each of the methods detected a gap.
In the figure, the green-shaded  region with dashed boundary indicates  where  Knabe's criterion  could detect a gap for system sizes $n = 3,\ldots,10$, i.e., where $\max_{n\leq 10}\delta_{\mathrm{Knabe}}(n)>0$; 
the blue-shaded region with dash-dotted boundary marks the same for the Gosset--Mozgunov bound. 
The results obtained with our LTI method (\cref{eq:primal_TI_Y}) are plotted in \Cref{fig:z3_r_s_deform} as individual data points. For each point on a grid in the $(r,s)$ plane  we plotted a marker indicating the   lowest $n$ for which a positive gap was detected  by the LTI SDP (triangle for $n=3$, square for 4, and pentagram for 5).

We observe in \Cref{fig:z3_r_s_deform} that  the LTI method for $n=5$ detects a gap in a larger region  than  Knabe's bound or the Gosset--Mozgunov bound do for $n=10$ (i.e.\, when we compare the methods' performance for roughly equal computational effort). 
Furthermore, the region detected by the $n=4$ LTI SDP contains all of the green shaded region (Knabe) and almost all of the blue shaded region (Gosset--Mozgunov).

\subsection{1D Glauber model}

 Glauber models are a family of quantum Hamiltonians related to thermal states of classical Hamiltonians. 
Under the assumption that these classical thermal states obey the detailed balance condition, they can be mapped to quantum states of the form
\begin{align}
    \ket{\psi} = \frac{1}{\sqrt{Z_N}}\sum_\sigma e^{-\beta H(\sigma)/2}\ket{\sigma}
\end{align}
where $H(\sigma)$ represents a class of classical kinetic Ising models.
For this class of Hamiltonians, the state above is found to be representable as a matrix product state, thus allowing for the construction of a parent Hamiltonian, i.e.\ a frustration-free Hamiltonian which has the state as its ground state~\cite{cirac_matrix_2021}.
Here we consider the family of parent Hamiltonians known as the 1D Glauber model~\cite{augusiak_quantum_2010}, which has the local term
\begin{align}
\label{eq:glauber_model}
    \begin{split}
        h (\gamma,\delta) =
        &\id
        -
        A(\gamma,\delta)X_i
        +
        B(\gamma,\delta)Z_{i-1}X_iZ_{i+1}
        \\
        &+\delta Z_{i-1}Z_{i+1}
        -
        \frac{\gamma(1+\delta)}{2}\left(Z_{i-1}Z_i
        +
        Z_iZ_{i+1}\right),
    \end{split}
\end{align}
where $X,Z$ are Pauli matrices, and $A$, $B$ are scalar coefficients given as
\begin{align}
\begin{split}
    A(\gamma,\delta) &= 
    \frac{1-\delta}{2}
    +
    \frac{(1+\delta)\sqrt{1-\gamma^2}}{2},
    \\
    B(\gamma,\delta) &=
    1-\delta-A(\gamma,\delta).
\end{split}
\end{align}
The model approaches criticality for $\gamma\to 1$, and the scaling  behavior of the gap in this regime allows for the study of dynamical exponents~\cite{augusiak_quantum_2010}.
Additionally, it can be verified that the eigenvectors of $h(\gamma,\delta)$ are only dependent on $\gamma$, and the eigenvalues are
\begin{align}
    0,0,0,0,2(1-\delta),2(1-\delta),2(1+\delta),2(1+\delta),
\end{align}
i.e., independent of $\gamma$.
Note that for calculating finite size criteria bounds, the local gap threshold $\epsilon_n$ (Sec.~\ref{sec:existing_methods}) corresponds to the gap of the projector Hamiltonian,  which is independent of  $\delta$. 

Finally, to compute the finite size criteria bounds for a 3-local Hamiltonian, we have to adapt the calculation of $\delta_{\mathrm{Knabe}}$ and $\delta_{\mathrm{GM}}$ using a two-step coarse-graining procedure.
We first block two $\mathrm{spin}-\frac{1}{2}$ particles of the original chain into one $\mathrm{spin}-\frac{3}{2}$ $(d=4)$ particle, 
\begin{align}
    \mathcal{I}=\{2i-1,2i\},
\end{align}
In the next step, we combine consecutive local terms of the original Hamiltonian (\cref{eq:glauber_model}) into one term,
\begin{align}
\label{eq:glauber_blocked_hamiltonian}
    \tilde{h}_{\mathcal{I},\mathcal{I}+1}&= h_{2i-1,2i,2i+1}+h_{2i,2i+1,2i+2},
\end{align}
which provides the expression for the nearest neighbor Hamiltonian $\tilde{h}$ on the new $\mathrm{spin}-\frac{3}{2}$ chain.
\begin{figure}[ht]
    \centering
    \includegraphics[clip, width=0.52\textwidth]{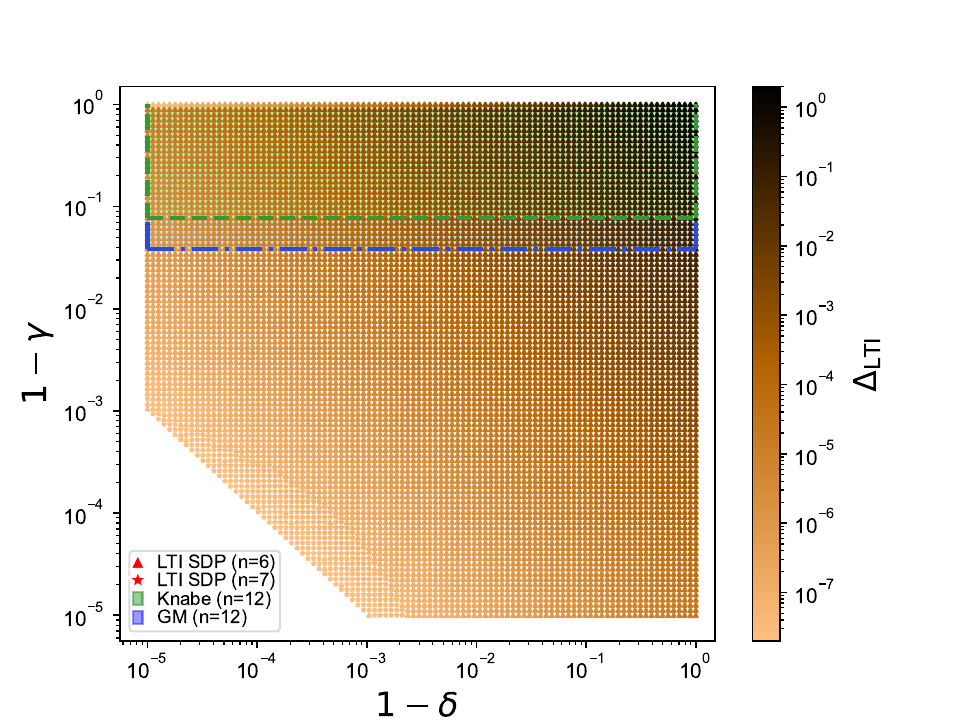}
    \caption{
    Gaps of the 1D Glauber Hamiltonian family.
    For a family of spin chain models parameterized by two parameters $\gamma,\delta$ (\cref{eq:glauber_model}) lower bounds on the gap were computed at different values of the parameters using different methods: 
    the Knabe and Gosset--Mozgunov finite size criteria (\cref{eq:knabe_bound} and \cref{eq:gosset_bound} respectively) computed for the coarse-grained  Hamiltonian $\tilde{h}$ (\cref{eq:glauber_blocked_hamiltonian}),
    and our  LTI SDP method \cref{eq:primal_TI_Y} computed for the original Hamiltonian. 
    The green shaded region (dashed border)  marks the region in which the Knabe method detected a gap ($\gamma\lesssim 0.92$),  the  blue shaded region (dash-dotted border) marks the region where  the Gosset--Mozgunov bound did ($\gamma\lesssim 0.96$). 
    The gaps detected by our LTI method are plotted as individual data points with marker colors corresponding to the size of the gap (see color bar).
    }
     \label{fig:kinetic-ising-plot}
\end{figure}
\Cref{fig:kinetic-ising-plot} shows the parameter region in which the finite size criteria detect a gap compared to the region in which the LTI SDP method does. 
For the range $\gamma,\delta\in[0,1)$, we computed the Knabe and Gosset--Mozgunov bounds for a system of 6 blocked spins ($n=12$ of the original spins) with interaction term $\tilde{h}$;
as well as the LTI SDP for $n=6$ 
with the original Hamiltonian from \cref{eq:glauber_model}.
For the LTI results the color of the marker at each point corresponds to the size of the detected gap (see color bar). 

We observe that  the LTI SDP allows us to detect a gap orders of magnitude closer to the critical value ($\gamma =1$) than what is possible using the two other finite size criteria.

\section{Conclusion and Outlook}
\label{sec:conclusion}
In this paper, we put forward a semidefinite programming hierarchy for certifying the spectral gap of frustration-free Hamiltonians.
The method relies on a well-known relation between the positivity of a quadratic operator in the Hamiltonian and the existence of a spectral gap. 
Each level $n$ in  our hierarchy searches for the maximal value $\delta_{\mathrm{LTI}}$ for which  the quadratic operator on the global system can be decomposed as a sum of \textit{$n$-local} positive terms.
Thus, it involves an $n$-body optimization problem that can be efficiently solved with semidefinite programming.

We proved that our method matches or outperforms  the most widely used finite-size criteria like the Knabe~\cite{knabe_energy_1988} or Gosset--Mozgunov~\cite{gosset_local_2016} bounds.
Extensive numerical benchmarks on 1D frustration-free models show that our method significantly outperforms 
the above criteria, when compared at the same computational cost. 
In particular, for the spectral gap of the AKLT chain, we provide bounds that  tightly match the best known estimates.
For other families of models we  demonstrated that our method detects gaps in a significantly larger parameter range than previous finite size criteria do,  especially in models close to criticality.

Interestingly, the dual formulation of our problem corresponds to the minimization of the first excited energy on marginals satisfying 
the so-called local translational invariance (LTI) property. 
Those marginals can be seen as a relaxation of the first excited reduced states. 
In this sense, our method adds to the recent advancements on SDP relaxations to estimate the ground state~\cite{kull_lower_2024,fawzi_entropy_2024}, and to compute its physical properties~\cite{wang_certifying_2024}.

An issue we did not resolve in this paper is the completeness of our hierarchy. 
Therefore, it remains open whether the hierarchy always detects a gap at some finite level when a model is indeed gapped,
and whether the certified gap always converges to the true gap as the level in the hierarchy is increased.
In this context, it is important to note the undecidability of the spectral gap problem for frustration-free Hamiltonians on 2D lattices \cite{cubitt_undecidability_2022,cubitt_undecidability_2015}. Coupling a complete hierarchy for lower bounds with a hierarchy of upper bounds obtained using variational methods would constitute a decision procedure for the spectral gap, violating the established undecidability results. We should therefore not expect to be able to prove completeness of our hierarchy in two or more spatial dimensions without further assumptions. Since the known 1D undecidability constructions \cite{bausch_undecidability_2020} are not frustration-free, the question of completeness of the hierarchy for one-dimensional frustration-free systems remains open.



We believe it is possible to adapt our method to the case of systems with open boundary conditions. 
It is known that finite-size criteria can be modified to take into account edge terms in the Hamiltonian~\cite{lemm_spectral_2019}, such as those in the Motzkin and Fredkin chains~\cite{movassagh_supercritical_2016,zhang_quantum_2016}.
We expect that, in a similar fashion, one can modify the LTI SDP to allow for the treatment of such systems as well. 

Finally, from its formulation, we believe that our method can allow the certification of gaps for systems in higher spatial dimensions. 
The generalization of our method to such systems is straightforward. 
The challenge in applying our method in this setting  lies in the ability to solve large scale SDPs.
For a matrix variable of a given size, the complexity solving an SDP  exceeds that of full exact diagonalization, which is required in order to certify the existence of a gap using finite size criteria.
Yet, from the results we presented for one-dimensional systems  it seems reasonable to expect that the system size for which one would need to solve the problem could be significantly smaller than the system sizes involved in  proving gaps using finite  size criteria in 2D (e.g.~\cite{lemm_aklt_2019, pomata_demonstrating_2020, lemm_existence_2020}).

\begin{acknowledgments}
We thank Mil\'an Rozm\'an for guiding us through the martingale literature. 
We thank Mingru Yang for enlightening discussions as well as for providing us with the DMRG result for the AKLT model. 
We thank Ignacio Cirac for stimulating discussions at the beginning of the project.
We thank Jin-Fu Chen for useful discussions about the Glauber model.
We thank Vedran Dunjko for useful discussions about undecidability.
We thank Anurag Anshu for bringing the coarse-graining method for higher-locality Hamiltonians to our attention.
I.K.\ and N.S.\ acknowledge support 
from the European Union’s Horizon 2020 research and innovation programme through Grant No.\ 863476 (ERC-CoG SEQUAM),
and the FWF Quantum Austria Funding Initiative project “Entanglement Order Parameters” 
(Grant DOI \href{https://doi.org/10.55776/P36305}{10.55776/P36305}), funded through the European Union – NextGenerationEU.
N.S.\ also acknowledges support by the 
the Austrian Science Fund FWF through Grant DOIs
\href{https://doi.org/10.55776/COE1}{10.55776/COE1} and
\href{https://doi.org/10.55776/F71}{10.55776/F71}.
The computational results presented have been achieved in part using the Austrian
Scientific Computing (ASC) infrastructure.
K.S.R.\ and J.T.\ acknowledge the support received from the European Union's Horizon Europe research and innovation programme through the ERC StG FINE-TEA-SQUAD (Grant No.~101040729). 
P.E.\ and J.T.\ acknowledge the support received by the Dutch National Growth Fund
(NGF), as part of the Quantum Delta NL programme. 
P.E.\ acknowledges the support received through the NWO-Quantum Technology
programme (Grant No.~NGF.1623.23.006).
P.E. further acknowledges funding by the Carl-Zeiss-Stiftung (CZS Center QPhoton).
This publication is part of the `Quantum Inspire - the Dutch Quantum Computer in the Cloud' project (with project number [NWA.1292.19.194]) of the NWA research program `Research on Routes by Consortia (ORC)', which is funded by the Netherlands Organization for Scientific Research (NWO).
Views and opinions expressed are however those of the authors only and do not necessarily reflect those of the funding institutions. Neither of the funding institutions can be held responsible for them.
Parts of this work were performed by using the compute
resources from the Academic Leiden Interdisciplinary Cluster Environment (ALICE) provided by Leiden University.
F.B. acknowledges financial support from the Alexander von Humboldt Foundation and support from the ICSC – Centro Nazionale di Ricerca in High Performance Computing, Big Data and Quantum Computing, funded by European Union – NextGenerationEU.

\end{acknowledgments}
\bibliographystyle{unsrtnat}
\bibliography{references}

\widetext
\appendix
 \section{The LTI condition is the only freedom in TI Hamiltonians}
 \label{app:LTI_proof}
In the main text, we used the property that if  $g_n$ and $q_n$ generate the same translation invariant operator 
on a spin chain with $m$ sites and periodic boundary conditions, i.e., if 
\begin{align}
    \sum_{i=1}^m \tau_i(q_{n}) = \sum_{i=1}^m \tau_i(g_{n}) ,
\end{align}
where $\tau_i$ is the translation operator such that $\tau_i(g_n)$ acts on sites $\{i,\ldots,i+n-1\}$,
then they are related as (\cref{eq:Y_condition} of main text):
\begin{align}
    q_{n} = g_{n}+A\x\id - \id\x A ,
\end{align}
for some $(n-1)$-body operator $A$.
We now prove this claim:  
 Let $X$ be an $n$-body operator and let $m\geq 2n-1$, then
 $\sum_{i=1}^m \tau_i(X)= 0$ holds  iff there exists an $(n-1)$-body operator $ A$ such that  $X = A \x \id- \id\x A$.

\textit{ Proof}: 
For the implication  $X = A \x \id- \id\x A$ $ \Rightarrow$ $\sum_{i=1}^m \tau_i(X)= 0$, notice that the terms $\tau_{i+1}(A \x \id)$ and $\tau_i(\id\x A)$  appear with opposite signs and thus the whole sum cancels telescopically, meaning that 
\begin{align}
   \sum_i \tau_i(A \x \id- \id\x A)=0.
\end{align}
 It remains to prove the other implication.
 To prove this, we take an orthonormal basis consisting of strings of single site operators $S_{i_1,i_2,\ldots,i_n}:= \sigma_{i_1}\x \sigma_{i_2}\x \ldots \sigma_{i_n}$ with $\sigma_0 $ proportional to the identity $ \id$.
We write out the coefficients of  $X$ in this basis,
 \begin{equation}
     X = \sum_{i_1,i_2,\ldots,i_n} X_{i_1,i_2,\ldots,i_n} S_{i_1,i_2,\ldots,i_n} .
 \end{equation}
 First consider the basis elements with $i_1\neq 0$, $i_n\neq 0$, and $i_2,\ldots,i_{n-1}$ arbitrary, and compute the trace of $S_{i_1,i_2,\ldots,i_n}\otimes\id_{n+1,\ldots, m}$ with $ \sum_j \tau_j(X)$. We see that only one term in the sum survives (the one with $j=1$):
 \begin{equation}
 0 = \textrm{Tr}\left(\sum^m_{j=1}\tau_j(X) S_{i_1,i_2,\ldots,i_n} \right) = X_{i_1,i_2,\ldots,i_n}, \;\; \text{for } i_1\neq0, i_n\neq0, \text{and }  i_2,\ldots,i_{n-1} = 0,\ldots, d^2-1 . 
 \end{equation}
 We thus see that the part of $X$ which has non trivial support on sites 1 and $n$ must vanish. 

Now let $n\geq k\geq2$ and let $i_j = 0$ for all $j<k$ and $i_k,i_n\neq 0$, and let $i_{k+1},\ldots,i_{n-1}$ be arbitrary. Taking the trace of $\sum_j\tau_j(X) $ with  $S_{0,0,\ldots,0,i_k,\ldots,i_n}\otimes\id_{n+1,\ldots, m}$ we get that
 \begin{equation}
 \label{condk}
    X_{i_k,\ldots,i_n,0,\ldots,0} + X_{0,i_k,\ldots,i_n,\ldots,0} + 
     \ldots + X_{0,\ldots,i_k,\ldots,i_n} = 0 . 
 \end{equation}
We will now use this relation between the coefficients of $X$ corresponding to basis elements with a non trivial $(n-k+1)$-body operator somewhere in the string (\cref{condk}) to show that the part of $X$ with non trivial support on  $n-k+1$ sites can be written as $A^{(n-k+1)}\otimes\id - \id\otimes A^{(n-k+1)}$ for some $A^{(n-k+1)}$.
Let $X^{(n-k+1)}$, ``the $(n-k+1)$-body part of $X$'',  be defined as 
\begin{equation}
 \label{Xk}
   X^{(n-k+1)} := \sum_{\substack{i_k\neq0,i_n\neq0, \\ i_{k+1},\ldots,i_{n-1}}}
    X_{i_k,\ldots,i_n,0,\ldots,0} S_{i_k,\ldots,i_n,0,\ldots,0}+  
    X_{0,i_k,\ldots,i_n,\ldots,0}
    S_{0,i_k,\ldots,i_n,\ldots,0}+
    \ldots + X_{0,\ldots,i_k,\ldots,i_n}
    S_{0,\ldots,i_k,\ldots,i_n}  . 
 \end{equation}
For $l= 1,\ldots,k$ define 
\begin{equation}
\label{T_contributions1}
T^{(n-k+1)}_l := \sum_{\substack{i_k\neq0,i_n\neq0, \\ i_{k+1},\ldots,i_{n-1}}}
X_{\underset{l-1\text{ times}}{0,\ldots\ldots,0},i_k,\ldots,i_n,\underset{k-l\text{ times}}{0,\ldots\ldots,0}} 
S_{i_k,\ldots,i_n}  \times d^{(k-1)/2},
\end{equation}
such that  
\begin{align}
    \label{Xk}
     X^{(n-k+1)} =  & T^{(n-k+1)}_1 \x\id \x \id \x \ldots\id 
      +\\ \nonumber
      & \id \x T^{(n-k+1)}_2 \x \id \x\ldots\id
      +\\ \nonumber
      \vdots
      \\ \nonumber
      &  \id \x \id \x \id \x\ldots T^{(n-k+1)}_k.
\end{align}
 Now for $j=1,\ldots,k-1$, we define    $T'^{(n-k+1)}_j = \sum_{l=1}^j T^{(n-k+1)}_l$, and  
\begin{align}
    A^{(n-k+1)} :=  
    & T'^{(n-k+1)}_1 \x \id  \x \id \x \ldots\id + 
    \\ \nonumber
    & \id\x T'^{(n-k+1)}_2\x\id \x \ldots\id +  
    \\ \nonumber
    & \vdots
    \\ \nonumber
    & \id \x\id\x\id \x\ldots T'^{(n-k+1)}_{k-1}.
\end{align}
\Cref{condk}  implies 
$T^{(n-k+1)}_k = -\sum_{l=1}^{k-1} T^{(n-k+1)}_l = - T'^{(n-k+1)}_{k-1}$
from which we obtain that 
\begin{equation}
X^{(n-k+1)} =  A^{(n-k+1)}\otimes\id - \id\otimes A^{(n-k+1)} .
\end{equation}
Since $X = \sum_{k=2}^n X^{(n-k+1)}$ (we have shown in the beginning that $X^{(n)} = 0$), we have  that  
$X  =  A \otimes\id - \id\otimes A  $, with 
$A = \sum_{k=2}^n A^{(n-k+1)}$ as claimed.

\section{Proof for inclusion of finite-size based methods}
\label{app:special_case}
We show that the finite size criteria bounds arise as a special case of the SDP~\cref{eq:primal_TI_Y}.
The operator $Q_n(\delta)$ which is a truncation of $H^2-\delta H$ is defined as,
\begin{align}
\label{eq:G_L_n}
\begin{split}
    Q_n(\delta_n) &= \sum_{i=1}^Nh_i^2
    +\sum_i\left(\{h_i,h_{i+1}\}+\{h_i,h_{i+2}\}+...+\{h_i,h_{i+n-2}\}\right)
    -\delta_n\sum_{i=1}^Nh_i,
\end{split}
\end{align}
where $\delta_n$ can take any value in the feasible set of the SDP in~\cref{eq:primal_TI_Y}.
For this we need to find a positive generating local term $q_n(\delta)\succeq 0$ such that,
\begin{align}
    Q_n(\delta_n) &= \sum_iq_{n,i}(\delta)\quad\text{ and }\quad q_{n,i}(\delta)\succeq 0.
\end{align}
In the following subsections we explicitly derive $q_n(\delta)$ for the bound $\delta$ given by the Knabe bound~\cite{knabe_energy_1988} (\cref{appendix:knabe_bound}) and  the Martingale bound~\cite{nachtergaele_spectral_1996} (\cref{appendix:martingale_bound}). 
For the Gosset--Mozgunov bound~\cite{gosset_local_2016} we prove that such a generating term exists without explicitly computing it (\cref{appendix:GM_bound}).

Following the assumptions of these methods, in the proof, we focus on the case of nearest neighbor Hamiltonians, and assume that the local terms are projectors, i.e. $h_i^2=h_i$.
\subsection{Knabe bound \cref{eq:knabe_bound} }
\label{appendix:knabe_bound}
The finite size criteria bounds are based on comparing the contribution of different overlapping and non-overlapping anticommutator terms $\{h_i,h_j\}$ in $H^2$.
The method is based on deriving an operator approximation for global operator $H^2$ (on the full system), in terms of the open boundary Hamiltonian on a finite system of $n$ sites.
We define the open boundary Hamiltonian as
\begin{align}
    h_{n,i}=\sum_{j=i}^{i+n-2}h_j.
\end{align}
where the index $i$ denotes the index of the starting site. 
Note that this Hamiltonian is supported on sites $i$ to $i+n-1$.
The open boundary Hamiltonian satisfies the following relation,
\begin{align}
\label{eq:gap_n_sites}
    h_{n,i}^2\succeq\epsilon_nh_{n,i},
\end{align}
where $\epsilon_n$ is a lower bound on the gap of $h_{n,i}$.
Then, the Knabe bound is given as
\begin{align}
\label{eq:knabe_bound_app}
    \Delta_m\geq\delta_{\mathrm{Knabe}}(n) = \frac{n-1}{n-2}\left(\epsilon_n-\frac{1}{n-1}\right).
\end{align}
It provides a direct relation between the finite size gap $\epsilon_n$ and the gap of system of size $m>n$.
To prove the inclusion of the Knabe bound, we require a positive operator decomposition for $Q_n(\delta_{\mathrm{Knabe}}(n))$.
The first step is to write the expansion of $h_{n,i}^2$, 
\begin{align}
    h_{n,i}^2 &= \left(h_{i}+...+h_{i+n-2}\right)^2
    \\
    &=h_i^2+...+h_{i+n-2}^2
    +\sum_{j=0}^{n-3}\{h_{i+j},h_{i+j+1}\}
    +\sum_{j=0}^{n-4}\{h_{i+j},h_{i+j+2}\}
    +...+\{h_i,h_{i+n-2}\}
    \\
    &=h_{n,i}
    +\sum_{j=0}^{n-3}\{h_{i+j},h_{i+j+1}\}
    +\sum_{j=0}^{n-4}\{h_{i+j},h_{i+j+2}\}
    +...+\{h_i,h_{i+n-2}\}
    \label{eq:h_n_sqr}
\end{align}
where in the last step, we used that the local terms are projectors $h_i^2=h_i$.
Summing on the left and right over all sites $i$,
\begin{align}
    \sum_{i=1}^Nh_{n,i}^2
    &=(n-1)H+(n-2)\sum_i\{h_i,h_{i+1}\}
    +(n-3)\sum_i\{h_i,h_{i+2}\}+...+\sum_i\{h_i,h_{i+n-2}\}.
\end{align}
Now we will derive a positive operator decomposition for  $Q_n(\delta_{\mathrm{Knabe}}(n))$.
First let us write the operator
\begin{align}
   Q_n(\delta_{\mathrm{Knabe}}(n)) &=H^2-\delta_{\mathrm{Knabe}}(n)H
   \\&=H
    +\sum_i\left(\{h_i,h_{i+1}\}+\{h_i,h_{i+2}\}+...+\{h_i,h_{i+n-2}\}\right)
    -\delta_{\mathrm{Knabe}}(n)H,
\end{align}
Substituting the Knabe bound \cref{eq:knabe_bound_app} into the above equation,
\begin{align}
   Q_n(\delta_{\mathrm{Knabe}}(n)) &= H
    +\sum_i\left(\{h_i,h_{i+1}\}+\{h_i,h_{i+2}\}+...+\{h_i,h_{i+n-2}\}\right)
    -\left(\frac{n-1}{n-2}\epsilon_n-\frac{1}{n-2}\right)H,
    \\
    &=
    \left(\frac{1-\epsilon_n}{n-2}\right)(n-1)H+
    \sum_i\{h_i,h_{i+1}\}+\{h_i,h_{i+2}\}+...+\{h_i,h_{i+n-2}\}
    \label{eq:Q_n_Knabe}
\end{align}
Now, we notice that $(n-1)H$ can be expanded as
\begin{align}
    (n-1)H = \sum_ih_i+h_{i+1}+...+h_{i+n-2},
\end{align}
Substituting this in \cref{eq:Q_n_Knabe},
\begin{align}
    Q_n(\delta_{\mathrm{Knabe}}(n)) &=
    \sum_i\left(\frac{1-\epsilon_n}{n-2}\right)(h_i+h_{i+1}+...+h_{i+n-2})+\{h_i,h_{i+1}\}+\{h_i,h_{i+2}\}+...+\{h_i,h_{i+n-2}\}.
\end{align}
In the above expression, $Q_n$ has a $n-$local decomposition.
However, we require a \textit{positive} $n-$local operator decomposition $q_n$, which can be obtained by rearranging the terms as allowed by the LTI freedom (\cref{eq:primal_TI_constraint}, App.~\ref{app:LTI_proof}).
For the anticommutators, we use the following replacements,
\begin{align}
    \{h_i,h_{i+1}\} &\rightarrow \frac{1}{n-2}\sum_{j=0}^{n-3}\{h_{i+j},h_{i+j+1}\}
    \\
    \{h_i,h_{i+2}\} &\rightarrow \frac{1}{n-3}\sum_{j=0}^{n-4}\{h_{i+j},h_{i+j+2}\}
    \\
    \vdots\nonumber
    \\
    \{h_i,h_{i+n-3}\} &\rightarrow \frac{1}{2}\left(\{h_{i},h_{i+2}\}+\{h_{i+1},h_{i+3}\}\right)
\end{align}
which when summed over $i$ on both sides, recover the same global operator.
For example, in the first replacement, the two sides are locally related by an operator $Y$ supported on $n-1$ sites, which satisfies the equation,
\begin{align}
    \{h_i,h_{i+1}\} +\id\x Y-Y\x\id = \frac{1}{n-2}\sum_{j=0}^{n-3}\{h_{i+j},h_{i+j+1}\}.
\end{align}
Using the above replacements, the new form for $Q_n$ is,
\begin{align}
    Q_n(\delta_{\mathrm{Knabe}}(n))
    &=\sum_i\left[\left(\frac{1-\epsilon_n}{n-2}\right)h_{n,i}
    +\frac{1}{n-2}\sum_{j=0}^{n-3}\{h_{i+j},h_{i+j+1}\}
    +\frac{1}{n-3}\sum_{j=0}^{n-4}\{h_{i+j},h_{i+j+2}\}+...+\{h_i,h_{i+n-2}\}\right]
    \\
    &:=\sum_iq_{n,i}(\delta_{\mathrm{Knabe}}(n)).
\end{align}
where above we chose a $n-$local form for $q_n$.
We need to show that 
\begin{align}
    q_{n,i}(\delta_{\mathrm{Knabe}}(n)) := 
    \left(\frac{1-\epsilon_n}{n-2}\right)h_{n,i}
    +\frac{1}{n-2}\sum_{j=0}^{n-3}\{h_{i+j},h_{i+j+1}\}
    +\frac{1}{n-3}\sum_{j=0}^{n-4}\{h_{i+j},h_{i+j+2}\}+...+\{h_i,h_{i+n-2}\}
\end{align}
is positive semidefinite, i.e. $q_{n,i}(\delta_{\mathrm{Knabe}}(n))\overset{!}{\succeq} 0$.
Using $h_{n,i}^2\succeq\epsilon_nh_{n,i}$,
\begin{align}
    q_{n,i}(\delta_{\mathrm{Knabe}}(n))
    &\succeq
    \frac{1}{n-2}(h_{n,i}-h_{n,i}^2)
    +\frac{1}{n-2}\sum_{j=0}^{n-3}\{h_{i+j},h_{i+j+1}\}
    +\frac{1}{n-3}\sum_{j=0}^{n-4}\{h_{i+j},h_{i+j+2}\}+...+\{h_i,h_{i+n-2}\}
\end{align}
Substituting $h_{n,i}^2$ from \cref{eq:h_n_sqr}, we obtain
\begin{align}
\begin{split}
    q_{n,i}(\delta_{\mathrm{Knabe}}(n))
    &
    \succeq
    \frac{1}{(n-3)(n-2)}\sum_{j=0}^{n-4}\{h_{i+j},h_{i+j+2}\}
    +\frac{2}{(n-4)(n-2)}\sum_{j=0}^{n-5}\{h_{i+j},h_{i+j+3}\}
    ...+\frac{n-3}{n-2}\{h_i,h_{i+n-2}\}
    \\
    &\succeq 0,
\end{split}
\end{align}
where the positivity follows because the RHS is sum of disjoint anticommutators.
Thus, we explicitly found a $n-$local positive operator decomposition for $[H^2]_n-\delta_{\mathrm{Knabe}}(n)H$, proving that $\delta_{\mathrm{Knabe}}(n)$ lies in the feasible set of the LTI SDP.

\subsection{Gosset--Mozgunov bound \cref{eq:gosset_bound}}
\label{appendix:GM_bound}
Define $B_{n,k}$ as in equation (8) in Ref.\ \cite{gosset_local_2016}: $B_{n,k} = \sum_{i=k}^{n-2+k}c_{i-k}h_{i,i+1}$, where $c_j$ are positive coefficients satisfying the conditions stated below.
Using equations (18) and (20) of Ref.\ \cite{gosset_local_2016}, which relate $\sum_k B_{n,k}^2$ and $[H^2]_n$, and using the expression for the Gosset--Mozgunov gap $\delta_{\mathrm{GM}}$ from equation (26) in same Ref., we obtain
\begin{align}
    [H^2]_n - \delta_{\mathrm{GM}}H &= \sum_i q_{n,i}(\delta_{\mathrm{GM}}) \\&=
    \frac{1}{\theta(1)}\left( \sum_iB_{n,i}^2 - \frac{\sum_j c_j}{n-1} \epsilon_n \sum_iB_{n,i} + \sum_{\substack{(j,k)\\n-2\geq d(j,k) \geq 2}} h_jh_k \left[\theta(1) - \theta(d(j,k))\right] \right) 
    \label{eq:gm_starting_op}
\end{align}
where
$\theta(x):=\sum_{k=0}^{n-x-2}c_kc_{k+x}$.
The coefficients $c_j$ in the deformed Hamiltonian $B_{n,i}$ satisfy the following properties
\begin{itemize}
    \item Positive: $c_j>0$ for $0\leq j\leq n-2$
    \item Non-decreasing upto midpoint: $c_j>c_{j-1}$ for $1\leq j\leq \frac{n-2}{2}$
    \item Symmetric about midpoint: $c_j = c_{(n-2)-j}$ for $0\leq j\leq\frac{n-2}{2}$
\end{itemize}
Given the above conditions, the coefficients $c_j$ satisfy the 1D autocorrelation lemma~\cite{gosset_local_2016} 
\begin{align}
    \theta(x)\geq \theta(x+1),\quad \forall\, x=0,1,...,n-3.
\end{align}
This implies that the term $\theta(1)-\theta(d)$ is non negative for any $n-2\geq d \geq 2$ and thus the third term in \cref{eq:gm_starting_op} is positive semidefinite.

To show inclusion of the GM bound, we want to obtain a local positive decomposition of 
the rest of \cref{eq:gm_starting_op}:
$\sum_i B_{n,i}^2 - \frac{\sum_j c_j}{n-1} \epsilon_n \sum_i B_{n,i}$.
As in the previous section, we are free to use the LTI freedom, i.e., add a term of the form $\id\otimes Y-Y\otimes\id$.
We will prove that a $Y$ exists such that 
$B_{n}^2 - \frac{\sum_j c_j}{n-1} \epsilon_n B_{n} + \id\otimes Y-Y\otimes\id \succeq 0$ without giving an explicit expression for it (unlike in the case of the Knabe bound in the previous section).

The proof of existence of such a $Y$ relies on the following fact: 
For an $n$-body operator $K$ the following SDPs are dual to each other and attain the same value.
\begin{align}
&\min_{\rho } \mathrm{Tr}K\rho  \\ \nonumber
&\text{s.t. }  \rho\succeq0  \\ \nonumber
&\Tr_1 \rho = \Tr_n \rho \\ \nonumber
&\Tr\rho =1
\end{align}
\begin{align}
&\max_{\epsilon,A} \epsilon \\ \nonumber
&\text{s.t. } K +\id\otimes Y - Y \otimes \id \succeq \epsilon \id   
\end{align}
Strong duality holds in this case because we have a feasible positive definite solution $\rho = \id / d \succ 0$. 
This implies that if  $K$ is non negative on  all LTI states ($\rho\succeq0$ such that $\Tr_1 \rho = \Tr_n \rho $)   then there exists a $Y$ such that  $K +\id\otimes Y - Y \otimes \id \succeq 0$.

It is therefore  sufficient to prove that 
\begin{equation}
\label{gm_to_prove}
\mathrm{Tr}\left[(B_{n}^2 - \frac{\sum_j c_j}{n-1} \epsilon_n B_n)\rho \right] \geq 0 \text{ for all } \rho\succeq0 \text{ such that } \Tr_1 \rho = \Tr_n \rho.
\end{equation}
The proof of \cref{gm_to_prove} is essentially the same as lemma 4 in \cite{gosset_local_2016}, the difference being that here $\rho$ is an $n$-body LTI state, whereas in \cite{gosset_local_2016} the state $\phi$ is a translation invariant eigenvector of the the $m$-site periodic chain Hamiltonian.

\textit{ Proof of \cref{gm_to_prove}}: Let $G^\perp$ be the projector to the complement of the ground space of $B_n$, and let $\hat{\rho} = G^\perp\rho G^\perp / \mathrm{Tr}G^\perp\rho$
\begin{align}
    \mathrm{Tr}\left[B_{n}^2 \rho \right] = \mathrm{Tr}\left[B_{n}^2 \hat{\rho} \right] \mathrm{Tr}G^\perp\rho
    \geq \mathrm{Tr}\left[B_{n} \hat{\rho} \right]
   \mathrm{Tr}\left[B_{n} \hat{\rho} \right]
   \mathrm{Tr}G^\perp\rho
\end{align}
where we used the non-negativity  of the variance of $B_n$ in $\hat{\rho}$.
Due to the LTI property of $\rho$ we have:
\begin{equation}
     \mathrm{Tr}\hat{\rho}h_i 
    = \frac{\mathrm{Tr}{\rho}G^\perp h_iG^\perp}{\mathrm{Tr}G^\perp\rho}
    =\frac{\mathrm{Tr}{\rho}h_i}{\mathrm{Tr}G^\perp\rho}
    = \frac{\mathrm{Tr}{\rho}h_j }{\mathrm{Tr}G^\perp\rho}
    =    \mathrm{Tr}\hat{\rho}h_j
\end{equation}
where the second equality is because for a pure state $\rho=\ket{\psi}\bra{\psi}$, we can decompose $\psi = G^\perp\psi + G\psi$, where $G$ is the ground state projector of $B_n$. Then $\bra{\psi}G^\perp hG^\perp\ket{\psi} = \bra{\psi}h\ket{\psi}$ because $hG\psi = 0$ ($B_n$ is frustration free). This implies the equality for mixed states.
The third equality is because $\rho $ is LTI.

Thus
\begin{align}
    \mathrm{Tr}\left[B_{n} \hat{\rho} \right] = 
    \sum_j c_j \mathrm{Tr} h_j \hat{\rho}  
    = \mathrm{Tr} h \hat{\rho} \sum_j c_j 
    = \frac{ \sum_j c_j}{n-1}\sum_i\mathrm{Tr} h_i \hat{\rho} = \frac{ \sum_j c_j}{n-1}\mathrm{Tr} H_n \hat{\rho} 
    \geq
    \frac{ \sum_j c_j}{n-1}\epsilon_n
\end{align}
So we get, 
\begin{align}
\mathrm{Tr}\left[B_{n}^2 \rho \right] \geq 
    \frac{ \sum_j c_j}{n-1}\epsilon_n
    \mathrm{Tr}\left[B_{n} \hat{\rho} \right]
   \mathrm{Tr}G^\perp\rho 
   = 
   \frac{ \sum_j c_j}{n-1}\epsilon_n
    \mathrm{Tr}\left[B_{n} \rho \right] 
\end{align}
as required.

\subsection{Martingale condition bound \cref{eq:martingale_bound}}
\label{appendix:martingale_bound}
We show that the martingale bound of the coarse-grained projector Hamiltonian $K = \sum_\mathcal{I}P_{\mathcal{I},\mathcal{I}+1}$ is contained in the feasible set of the LTI SDP for gap of Hamiltonian $K$.

We modify the original $k-$local Hamiltonian into the coarse-grained projector Hamiltonian $K$ through the following steps.
First, each site $\mathcal{I}$ of the coarse grained system is comprised of $l\geq k-1$ sites of the original system
\begin{align}
    &\mathcal{I}=\{(i-1)l+1,...,il\},
\end{align}
And the coarse grained Hamiltonian is defined as
\begin{align}
&\tilde{h}_{\mathcal{I},\mathcal{I}+1} = \sum\limits_{j=(\mathcal{I}-1)l+1}^{(\mathcal{I}+1)l-k+1}h_{j,...,j+k-1}.
\end{align}
The locality of the new Hamiltonian $\tilde{h}_{\mathcal{I},\mathcal{I}+1}$ is $2l$-local in the original system. 
Next we define the corresponding projector Hamiltonian $K$,
\begin{align}
K = \sum_\mathcal{I}P_{\mathcal{I},\mathcal{I}+1},
\\
\tilde{h}_{\mathcal{I},\mathcal{I}+1}\succeq\epsilon_{2l}P_{\mathcal{I},\mathcal{I}+1}.
\end{align}
To prove inclusion of the martingale bound, we will mainly use the inequality (Lemma 6.3(2) of~\cite{fannes_finitely_1992})
\begin{equation}
\label{eq:anticomm_bd}
    \{P_{\mathcal{I},\mathcal{I}+1},P_{\mathcal{I}+1,\mathcal{I}+2}\} \succeq
    - \eta_l(P_{\mathcal{I},\mathcal{I}+1}+P_{\mathcal{I}+1,\mathcal{I}+2}), 
\end{equation}
where 
\begin{align}
    \eta_l = \vert\vert(\id-G_{1,\ldots,2l})  (\id-G_{l+1,\ldots,3l})  - (\id-G_{1,\ldots,3l})  \vert\vert,
\end{align}
and $G_{i,i+1,...,i+m-1}$ denotes the projector onto the ground space of the original Hamiltonian $h$ on total $m$ sites from indices $i$ to $i+m-1$.

Then, $K^2$ can be lower bounded as
\begin{align}
     K^2 \geq & K + \sum_\mathcal{I} \{P_{\mathcal{I},\mathcal{I}+1},P_{\mathcal{I}+1,\mathcal{I}+2}\} \\
        \geq & K - \eta_l \sum_\mathcal{I}(P_{\mathcal{I},\mathcal{I}+1}+P_{\mathcal{I}+1,\mathcal{I}+2})  = (1-2\eta_l) K ,
\end{align}
which means that $\Delta$, the gap of Hamiltonian $K$, satisfies
\begin{align}
    \Delta\geq 1-2\eta_l .
\end{align}

To prove that the above bound is included in the feasible set, we find a positive and local decomposition for $Q_{3l}(1-2\eta_l)=[K^2]_{3l}-(1-2\eta_l)K$, where the subscript $3l$ refers to truncation size in the original system
\begin{align}
Q_{3l}(1-2\eta)&=[K^2]_{3l}-(1-2\eta_l)K
\\
&:=\sum_iq_{3l,i}(1-2\eta_l)
\end{align}
such that $q_{3l,i}\succeq 0$.
We start by expanding the LHS,
\begin{align}
    Q_{3l}(1-2\eta_l)
    &=
    K+\sum_{\mathcal{I}}\{P_\mathcal{I},P_{\mathcal{I}+1}\}
    -(1-2\eta_l)K
    \\
    &=
    \sum_i2\eta_l P_{\mathcal{I}}+\{P_{\mathcal{I}},P_{\mathcal{I}+1}\}
    \\
    &=
    \sum_i\eta_l (P_{\mathcal{I}}+P_{\mathcal{I}+1})+\{P_{\mathcal{I}},P_{\mathcal{I}+1}\}
\end{align}
Defining the local term to be
\begin{align}
    q_{3l,i}(1-2\eta_l) := \eta_l(P_{\mathcal{I}}+P_{\mathcal{I}+1})+\{P_{\mathcal{I}},P_{\mathcal{I}+1}\}
\end{align}
Using the relation \cref{eq:anticomm_bd}, we obtain that
\begin{align}
q_{3l,i}(1-2\eta_l)\succeq 0.
\end{align}
\section{Optimizing the interaction term using the SDP gradient formula}
\label{app:SDP_gradient}
It is known that the gradient of the optimal objective of an  SDP  with respect to the problem data can be obtained from the primal-dual pair of variables attaining the optimal solution \cite{reehorst_navigator_2021}.

First of all, we see that upon applying the SDP gradient formula (\cite{reehorst_navigator_2021}, Section 4.2)  to the exact SDP formulation of the gap problem \cref{eq:exactPrimal}, we recover the result of the Hellmann--Feynman theorem:
Let $H$ be the full system Hamiltonian with ground-state  energy zero, and let $P_0$ be the projector to the global ground state.
Assume $H$ depends on a parameter $x$
as $H(x) = H_0 + (\id-P_0)A(x)(\id-P_0)$, where $A$ is some operator depending smoothly on $x$, such that $H(x)$ has the same ground state as $H_0$ in a neighborhood of $x_0$.

The  SDP gradient formula tells us that the derivative of the objective function (in our case the exact gap) with respect to the parameter $x$  is given by the \emph{partial derivative} of the Lagrangian at the primal-dual optimal point. 
In the same spirit as the Hellmann–Feynman theorem---only the explicit dependence on $x$ contributes to the gradient.
Applying the SDP gradient formula to the SDP \cref{eq:exactPrimal}, where the corresponding Lagrangian is $\mathcal{L}(\rho,\delta) = \delta + \mathrm{Tr}(\rho (H^2-\delta H))$, and the  primal-dual optimum is ($\rho^{\star} = |\psi_1\rangle \langle \psi_1|/\Delta,\delta^{\star}=\Delta$),  we get 
\begin{align}
    \partial \Delta / \partial x  = 
    \Tr(\rho^\star \frac{\partial( H^2-\Delta H) }{\partial x }) =
   \langle \psi_1| \frac{ \partial H }{\partial x } |\psi_1\rangle, 
\end{align}
where $\psi_1$ is the first excited state with energy $\Delta$.

We can use the same gradient formula for the LTI SDP in \cref{eq:primal_TI_Y}.
This can be useful for example if we  are given a parent Hamiltonian with interaction term $h_0$ and we wish to find a different Hamiltonian with the same ground state but with a bigger gap. 
We can use the solution of the  SDP to perform a gradient search over the interaction term $h$ to maximize the gap (while maintaining frustration freeness).

Let $p_0$ be the projector to the ground space of $h_0$.
Any interaction term of the form $h(X):= h_0 + (\id-p_0)X(\id-p_0)$ which is still positive semidefinite and frustration free will have the same global ground state. 
We can use the SDP gradient formula  to differentiate the gap which the SDP detects with respect to  $X$.
Of course we can  always increase the gap arbitrarily by choosing $X=\mathrm{const.}\times\id$, so in addition to the positivity constraint $h(X)\succeq 0$ we should impose some constraint on $X$, e.g.\ $||X||_2\leq 1$. 
Both of those constraints are easily differentiable and so we could perform constrained gradient optimization.

Let us now consider the case of the LTI gap detection SDP \cref{eq:primal_TI_Y}. 
Applying the SDP gradient formula (\cite{reehorst_navigator_2021}, Section 4.2) we see that the gradient of the gap  is given by
\begin{align}
   \partial (\Delta_{\text{LTI}}(n))/ \partial X_{\alpha,\beta} 
   = 
   \Tr(\rho^{\star} \frac{\partial g_n(X,\delta^{\star})} {\partial X_{\alpha,\beta}}  )   
\end{align}
where $g_n(X,\delta ) = -\delta h_1(X) + h_1(X)^2 + 
    \sum_{j=2}^{n-1} \{h_1(X),h_j(X)\}   $.

\section{Certifiable formulation of the LTI SDP}
\label{app:certify_solution}
To regard the solution of the SDP as a  certificate or proof  of a gap $\geq \delta$ it is crucial that we have a handle on numerical precision errors. We now explain how the LTI SDP needs to be modified in order to control for such errors.
Recall the LTI SDP:
\begin{align}
\label{eq:LTI_SDP}
\Delta_{\mathrm{LTI}}(n) &= \max_{\delta,Y} \; \delta 
    \\ \nonumber
    \text{s.t. } & \quad  g_n(\delta)+\id\otimes Y - Y\otimes\id\succeq 0. \\  \nonumber 
    &  \quad g_{n} (\delta) :=\left(h_1^2+\sum\limits_{1<j<1+n-k}\{h_1,h_j\}-\delta h_1\right).
\end{align}
We first demonstrate the problem of regarding a finite-precision solution as a proof.
Typically the solver produces a solution $(\delta,Y)$ for which the constraint in \cref{eq:LTI_SDP} is positive semidefinite only up to some finite precision $\epsilon>0$ : 
\begin{equation}
\label{eq:finite_precisoin_q}
     q_n(\delta) := g_n(\delta)+\id\otimes Y - Y\otimes\id\succeq -\epsilon, 
\end{equation}
(for MOSEK, which is the solver we use, $\epsilon\approx  1e-8$). 
Such a solution cannot be used to prove a gap in the thermodynamic limit (or for large finite systems) because 
from $q_n$ as in \cref{eq:finite_precisoin_q} we can only infer
\begin{align}
    \sum^N_i   q_{n,i} \succeq -N\epsilon .
\end{align}
Therefore, to produce rigorous certificates out of finite-precision solutions, we need to find a way to deal with the solver's inaccuracy.



To make our solution certifiable we slightly modify the LTI SDP:  First we note that all the terms in $g_n$ annihilate any ground state of the $n$-body chain with open boundary conditions, as every ground state is annihilated by each individual term $h_i$. 
We choose to make the $Y$ terms in $q_n$ such that they also annihilate every ground state.
To ensure this, we construct the projector to the orthogonal complement of the ground space of the $n-1$ site chain, which we denote as $P_{n-1}^{\perp}$, and replace $Y$ with  $P_{n-1}^{\perp} Y P_{n-1}^{\perp}$ in \cref{eq:LTI_SDP}.  This ensures that the whole operator $q_n$ annihilates any ground state of the open chain. 
The optimal value $\delta$ found by the solver might now be lower because we restricted the set of allowed $Y$s, but this is acceptable, since it still represents a lower bound on the gap.
The projectors $P_{n}$ can be either computed by exact diagonalization  of the $n$-site Hamiltonian (we are always in the regime when this is possible as it is a subroutine needed to solve the SDP), or be given exactly by the matrix product state (MPS) representation of the ground state:\ the MPS with open bond indices spans the ground space of the open chain. The latter is the case  for all the models we treated in this paper.
The new form for the certifiable LTI SDP is then 
\begin{align}
\label{eq:LTI_SDP_proj}
\Delta^{\textbf{cert}}_{\mathrm{LTI}}(n) &= \max_{\delta,Y} \; \delta \\ 
    \text{s.t. } & \quad  g_n(\delta)+\id\otimes P_{n-1}^{\perp}YP_{n-1}^{\perp} - P_{n-1}^{\perp}YP_{n-1}^{\perp}\otimes\id\succeq 0. \\  
    &  \quad g_{n,i} (\delta) :=\left(h_i^2+\sum\limits_{i<j<i+n-k}\{h_i,h_j\}-\delta h_i\right)
\end{align}

With the above modification we have ensured that, independently of the value of $\delta$, the kernel of $q_n(\delta)$ contains all the ground states of the open chain Hamiltonian. 
Other zero (or small negative) eigenvalues of the optimal  $q_n(\Delta^{\textbf{cert}}_{\mathrm{LTI}}(n))$ found by the solver arise from the maximization over $\delta$ as part of the SDP solution. 
We can then check the minimal eigenvalue of  $q_n$ outside of its kernel and try to make it positive by modifying the gap estimate $\Delta^{\textbf{cert}}_{\mathrm{LTI}}(n)$ produced by the solver. 
For example, we can perform a line search over $\eta\in(0,1]$, halting the search when 
$\lambda_{\mathrm{min}}(P_{n}^{\perp} q_n(\Delta^{\textbf{cert}}_{\mathrm{LTI}}(n)-\eta)P_{n}^{\perp})$ is bigger than some positive threshold. 
In practice it is usually sufficient to  introduce a correction on the order of $\epsilon / \gamma_2(h_i)$, where $\gamma_2(h_i)$ is the gap of the interaction term, to get a strictly positive $q_n$.

\end{document}